\documentclass[aps,prc,superscriptaddress,showpacs,amsmath,amssym,
nofootinbib,twocolumn, aastex]{revtex4-1}

\usepackage{array}
\usepackage{graphicx}
\usepackage{epstopdf}
\usepackage{xcolor}
\usepackage{bm}
\usepackage{cmap}
\usepackage{nicefrac}
\usepackage{xfrac}
\usepackage{cleveref}
\usepackage{booktabs}
\usepackage{tabularx}

\begin{document}

\title[Neutrino luminosities and heat capacities of neutron stars]{Neutrino luminosities and heat capacities of neutron stars in analytic form}

\author{D. D. Ofengeim}
\affiliation{Ioffe Institute, 26 Politekhnicheskaya,
        St.~Petersburg 194021, Russia}
\author{M. Fortin}
    \affiliation{N. Copernicus Astronomical Center, Bartycka 18, 00-716 Warsaw, Poland}
\author{P. Haensel}
    \affiliation{N. Copernicus Astronomical Center, Bartycka 18, 00-716 Warsaw, Poland}
\author{D.\ G.\ Yakovlev}
    \affiliation{Ioffe Institute, 26 Politekhnicheskaya,
        St.~Petersburg 194021, Russia}
\author{J. L. Zdunik}
    \affiliation{N. Copernicus Astronomical Center, Bartycka 18, 00-716 Warsaw, Poland}

\date{\today}

\begin{abstract}
We derive analytic approximations for the neutrino luminosities and the heat
capacities of neutron stars with isothernal nucleon cores as
functions of the mass and radius of stars.  The neutrino luminosities
are approximated for the three basic neutrino emission mechanisms,
and the heat capacities for the five basic combinations of {the} partial
heat capacities. The approximations  are valid for for a wide class of 
equations of state of dense nucleonmatter. The results significantly simplify  
the theoretical interpretation of observations of cooling neutron stars as well as of quasistationary thermal states of neutron stars in X-ray
transients. For illustration, we present an
analysis of the neutrino cooling functions of nine isolated neutron
stars taking into account the effects of their magnetic fields and
of the presence of light elements in their heat blanketing
envelopes. These results allow one to investigate the superfluid
properties of neutron star cores.
\end{abstract}

\maketitle

\newcommand{\Phiwave}{\widetilde{\Phi}}
\newcommand{\const}{\mathrm{const}}
\newcommand{\Tg}{\widetilde{T}}
\newcommand{\Ts}{T_{s}}
\newcommand{\ddo}[1]{\textcolor{blue}{#1}}
\newcommand{\dg}[1]{\textcolor{red}{#1}}  

\section{Introduction}
\label{s:I}

Modeling {the} thermal evolution of isolated neutron stars as well as neutron stars
in X-ray transients (XRTs) and comparing the results with observations
is known to give a viable method to probe superdense
matter in neutron star interiors (e.g., Refs.\
\cite{YakHaens2003,YLH2003,YakPeth2004,PLPS2009,Pot2015} and
references therein).

The cooling of isolated neutron stars proceeds through the neutrino
emission from the entire stellar volume and through the heat
conduction to the surface and subsequent thermal surface emission of
photons. Young and middle-aged isolated neutron stars (of age $t
\lesssim 10^5$ yr) cool mostly via neutrino emission from their
internal layers. Initially, at $t \sim$ 10--100 yr after their
birth, neutron stars are nonisothermal inside; their cores stay
colder than the crust because of {the} stronger neutrino emission from the
cores and {the} lower thermal conduction in the crust (e.g., Ref.\
\cite{Yak2001} and references therein). However, when the initial
period of {internal} thermal relaxation is over, the interior of
the star becomes nearly isothermal. It has the same internal
temperature $\Tg$ which gradually decreases with time as the star
cools. {With} account for the effects of General Relativity, $\Tg$
should be the temperature redshifted for a distant observer (not the
local temperature of the matter, see below). A substantial
temperature gradient remains only in the outer heat blanketing
envelope of the star (not deeper than a few hundred meters under the
surface \cite{GPE1983}). {The} cooling of isolated neutron stars with {an}
isothermal interior at the neutrino cooling stage is governed (e.g.,
Ref. \cite{YakPeth2004}) by the ratio
$L_{\nu}^\infty(\Tg)/C({\Tg})$, where $L_\nu^\infty(\Tg)$ is the
neutrino luminosity of the star (the superscript $\infty$ means that
it is redshifted for a distant observer) and $C(\Tg)$ is the heat
capacity. Both, $L_\nu^\infty(\Tg)$ and  $C(\Tg)$, are mainly
determined by the neutron star core and can be calculated as
\begin{equation}
\label{eq:L-start}
L_{\nu}^\infty(\Tg) = \int_0^{R_{\rm core}} Q( \rho, T)
\frac{ \exp(2\Phi) \,4\pi r^2 \,{\rm d}r}{\sqrt{1 - 2 G m/(r c^2)}},
\end{equation}
\begin{equation}
\label{eq:C-start}
C_{\rm core}(\Tg) = \int_0^{R_{\rm core}} c_{\text{core}}( \rho,T)\,
\frac{4\pi r^2\, {\rm d}r}{\sqrt{1 - 2 G m /(r c^2)}}.
\end{equation}
Here, $Q(\rho,T)$ and $c_{\rm core}(T,\rho)$ are the neutrino
emissivity [erg~cm$^{-3}$~s$^{-1}$] and the specific heat capacity
[erg~cm$^{-3}$~K$^{-1}$] in the core, respectively; $\rho$ is the
mass (energy) density, $T=\Tg\,\exp(-\Phi)$ is {the} local temperature
in the stellar matter, $R_{\rm core}$ is the core radius, $m = m(r)$
is the gravitational mass inside {a} sphere of circumferential
radius $r$, and $\Phi=\Phi(r)$ is the metric function determined by
the equation (e.g., chapter 6 of Ref.~\cite{HPY2007})
\begin{equation}
\label{eq:dPhi}
\frac{{\rm d}\Phi}{{\rm d}r} = -\frac{1}{ P + \rho c^2}\,\frac{{\rm d}P}{{\rm d}r},
\end{equation}
with $P$ being the pressure.

Isolated neutron stars with $10-100$ yr $ \lesssim t \lesssim 10^5$~yr
 have isothermal interiors and cool mainly from inside via
neutrino emission (e.g., Refs.\
\cite{YakPeth2004,PLPS2009,Pot2015}). To study their cooling, one
needs $L_{\nu}^\infty(\Tg)$ and $C_{\rm core}(\Tg)$. Later, at $t
\gtrsim 10^5$ yr, the neutrino emission weakens and the stars cool
mostly via heat conduction to the surface and the surface thermal
emission. In order to investigate their cooling at this stage, one
needs $C_{\rm core}(\Tg)$.

Cooling theory is also used to analyze thermal states of neutron
stars in XRTs (e.g., Ref.\ \cite{YakPeth2004}), which are compact
binaries containing a neutron star and a low-mass star. These old
systems show active periods (days, weeks, or months) of accretion from
the low-mass companion to the neutron star through an accretion
disk. The active periods are superimposed by quiescent periods
(months or years) when the accretion stops.

During an active period, a huge amount of gravitational energy is released
when the matter falls {onto} the neutron star surface. This makes
XRTs bright X-ray sources. In addition, the accreted matter burns in
thermonuclear reactions in the surface layers which intensifies the
surface X-ray emission. The ashes of {the} thermonuclear burning are
further compressed under the weight of the newly {accreted} material
and undergo nuclear transformations (pycnonuclear reactions,
electron captures, and neutron emission or absorption) producing the
deep heating of the neutron star's crust
\cite{HZ1990,HZ2003,HZ2008,Brown1998} with an energy release of
about 1--2 MeV per accreted nucleon. This heat, whose power is
proportional to the accretion rate, is mainly conducted into the
core, warms it up and is radiated by neutrinos from there. As
episodes of heating due to the accretion and subsequent quiescence
with neutrino cooling proceed, the neutron star interior reaches a
state of thermal quasiequilibrium. Then the source operates in a
quasistationary regime and the interior of the star remains
isothermal because of {the} large internal thermal conduction. The
internal temperature $\Tg$ does not show noticeable variations since
the star is thermally inertial. The star stays thermally balanced
being heated during accretion episodes but cooled during quiescent
states. In quiescence, the violent processes of surface energy
release stop, and the surface temperature drops, but its decrease is
limited because the star is still warm inside. Thus the star can
produce an intense and detectable quiescent surface emission
\cite{Brown1998}. The internal temperature $\Tg$ of this star is
determined by its neutrino luminosity $L_\nu^\infty(\Tg)$ and by the
deep crustal heating power (e.g., Ref.\ \cite{YakPeth2004}).

If accretion episodes are long or intense, the crustal heating can
drive the crust out of the thermal balance with the core. This
balance is restored later, during subsequent quiescent states.
Observations of the relaxation of such neutron stars combined with
observations of an accretion outburst allow one to estimate a lower
limit to the heat capacity of the neutron star core
\cite{2017Cumming}.

Therefore, the thermal evolution of isolated and accreting neutron
stars is largely regulated by $L_\nu^\infty(\Tg)$ and $C(\Tg)$ which are
the quantities we analyze below. While modeling the thermal
evolution, it is often time consuming to calculate $L_\nu^\infty(\Tg)$ and
$C(\Tg)$ directly from Eqs. (\ref{eq:L-start}) and
(\ref{eq:C-start}). We will obtain convenient analytic fits which
considerably simplify the calculations and interpretation of
observational data. Note that some fits to
$L_\nu^\infty(\Tg)/C(\Tg)$ were obtained earlier in
Refs.~\cite{Yak2011,Of2015-NS}; they are in good agreement with the
present results but less complete. Preliminary results of this
investigation have been presented in Ref.\ \cite{2016OFEN}.

\section{Basic cases}
\label{s:2}

The quantities $L_\nu^\infty(\Tg)$ and $C(\Tg)$ in question are determined
(i) by an equation of state (EOS) of superdense matter {for the} neutron
star core and {an} appropriate neutron star model ({for given} mass $M$
and radius $R$) and (ii) by the neutrino emissivity $Q(\rho,T)$ and {the}
specific heat capacity $c_{\rm core}(\rho,T)$ in the core. {To be specific,}
 we will restrict ourselves to the case in which the core
consists of strongly degenerate neutrons ($n$), protons ($p$),
electrons ($e$), and muons ($\mu$), and consider a number of
different EOSs of $npe\mu$ matter. As a rule, neutrons are the most
abundant particles (e.g., Ref.\ \cite{HPY2007}). Neutrons and
protons constitute a strongly interacting Fermi liquid; in the {cores}
of massive neutron stars, these particles become mildly
relativistic. As for electrons and muons, they constitute a weakly
interacting Fermi gas; the electrons are ultrarelativistic while
the muons are typically mildly relativistic.

Even in this simplest case, the problem of calculating $L_\nu^\infty(\Tg)$
and $C(\Tg)$ is strongly complicated by {possible superfluidities of
neutrons and protons}. The critical temperatures for such
superfluidities are difficult to calculate exactly (e.g., Ref.\
\cite{PageReview2013} and references therein). Neutron and/or
proton superfluids affect $L_\nu^\infty(\Tg)$ and $C(\Tg)$
because of the onset of energy gaps in {the} nucleon dispersion relations.
Strong superfluidity exponentially suppresses {the} neutrino emissivities
for the neutrino processes involving superfluid particles and the
partial heat capacities for superfluid particles (as reviewed in
Ref. \cite{Yak2001}). In addition, it creates a new specific
mechanism of neutrino-pair emission due to {the} Cooper pairing of
nucleons (see Refs. \cite{1976FRS,2006LP,PLPS2009,Pot2015} and
references therein). If {the} superfluid critical temperatures {were} known,
{one could in principle numerically compute  $L_\nu^\infty(\Tg)$ and $C(\Tg)$; this could be a good project for the future.}

Here, we will follow the strategy used previously in
Refs.\ \cite{Weisskopf_etal11,Yak2011,SY2015,Kloch2015,Of2015-NS}
for analyzing the observations of some selected neutron stars.
First, we calculate {the} partial contributions to $L_\nu^\infty(\Tg)$ and
$C(\Tg)$ for {the} most important cases neglecting the effects of
superfluidity. We will approximate these contributions by analytic
expressions and use them as the basis {for the} neutrino luminosities and
heat capacities. The actual $L_\nu(\Tg)$ and $C(\Tg)$ can be
expressed as sums over the contributions from the various species.
However if one of them is strongly superfluid then the contribution
is strongly reduced (either completely or by an unknwon reduction factor).
By comparing the theory with observations using this procedure one can try to
constrain the superfluid properties of neutron stars.

\begin{table}
    \begin{center}
        \caption{\label{tab:nucases}
            Three basic neutrino processes}
        \renewcommand{\arraystretch}{1.4}
        \begin{tabular}{lccc}
            \hline\hline
            Case  & $L_\nu^\infty(\Tg)$   & Neutrino process & Superfluids (SF) \\
            \hline
            DU &  $\Tg^6$   & $n \to p\ell\widetilde{\nu}_l,~~p\ell \to n \nu_l $  & None    \\
            MU & $\Tg^8$   &  $nN \to pN \ell\widetilde{\nu}_l,~~pN\ell \to n N\nu_l $  & None \\
            $nn$ & $\Tg^8$   & $nn \to nn \nu \widetilde{\nu} $   & Strong $p$ SF   \\
            \hline\hline
        \end{tabular}
        \begin{tabular}{ll}
        $N$ = $n$ or $p$; \quad & $\ell$= $e$ or $\mu$
        \end{tabular}
    \end{center}
\end{table}

Our three basic cases for $L_\nu^\infty(\Tg)$ are presented in Table
\ref{tab:nucases}, where $N$ denotes a nucleon, and $\ell$ is either
an electron or a muon. The first column labels the cases, the second
one shows the temperature dependence of the corresponding neutrino
luminosities $L_\nu^\infty(\Tg)$, the third presents the leading
neutrino reaction and the last column indicates {the} superfluid state of
the neutron star core at which the given  $L_\nu^\infty(\Tg)$ is the
leading one. The first case is the most powerful direct Urca (DU)
neutrino cooling process \cite{LPPH1991} (a sequence of neutron
decay and inverse reactions producing an electron or a muon neutrino pair).
The DU process can be open only in the inner cores of massive
neutron stars with those EOSs which predict a sufficiently large
fraction of protons. The second case corresponds to a less powerful
neutrino cooling due to the modified Urca (MU) process (e.g. Ref.\
\cite{Yak2001} and references therein). This process is considered
{as standard} in not very massive nonsuperfluid neutron stars.
Finally, the last case $nn$ is for the neutrino-pair bremsstrahlung
due to neutron-neutron collisions (reviewed, e.g., in Ref.\
\cite{Yak2001}). In a nonsuperfluid star it is weaker than the MU, but
if protons are strongly superfluid, MU and DU processes are greatly suppressed and the neutrino-pair bremsstrahlung becomes the leading neutrino
cooling process.

\begin{table}
    \begin{center}
        \caption{\label{tab:capcases}
            Five basic heat capacities}
        \renewcommand{\arraystretch}{1.4}
        \begin{tabular}{lc}
            \hline\hline
            Case  & $C_{\rm core}(\Tg)$   \\
            \hline
            $n$ &  $C_n$    \\
            $p$ &  $C_p$    \\
            $\ell$ & $C_{\ell}=C_{e}+C_{\mu}$   \\
            $tot$ & $C_{tot}=C_{n}+C_{p}+C_{e}+C_{\mu}$   \\
            $n\ell$ & $C_{n\ell}=C_{n}+C_{e}+C_{\mu}$   \\
            \hline\hline
        \end{tabular}
    \end{center}
\end{table}

Table \ref{tab:capcases} presents {the} five basic cases of partial
(nonsuperfluid) heat capacities $C_{\rm core}(\Tg)$ {in} neutron star
cores. Case $n$ refers to the heat capacity of neutrons; case $p$ to
the heat capacity of protons; case $\ell$ to the sum of heat
capacities of electrons and muons, case $tot$ to the sum of the heat
capacities of all constituents of the matter, and case $n\ell$ to
the heat capacity of neutrons, electrons and muons. Case $tot$ gives
the total heat capacity of a nonsuperfluid core. Other cases can be
useful in the presence of superfluidity of nucleons. For instance,
case $n\ell$ corresponds to a strong superfluidity of protons, while
case $\ell$ to a strong superfluidity of neutrons and protons. All
basic heat capacities are proportional to $\Tg$ because of the strong
degeneracy of all fermions in the neutron star cores (e.g., Ref.\
\cite{Yak2001}).

\subsection{Neutrino luminosities}
\label{s:Lnu}

Let us outline the neutrino emissivities $Q(\rho,T)$ for the three
basic neutrino emission cases in Table \ref{tab:nucases} (e.g.,
Ref.\ \cite{Yak2001}).

For the MU process,
\begin{equation}
\label{eq:Qmu} Q_{\rm MU} = Q_{\rm MU\,0} \left( \frac{n_{p}}{n_0}
\right)^{1/3} T_9^8\, \Omega\left( n_{n}, n_{p}, n_{e}, n_{\mu}
\right),
\end{equation}
where $n_{\alpha}$ is {the} number density of particles
$\alpha=n,\ldots,\mu$; $n_0 = 0.16$~fm$^{-3}$ is the standard number
density of nucleons in saturated nuclear matter, $T_9$ is the local
temperature $T$ the expressed in $10^9$~K and $\Omega\sim 1$ is a
dimensionless factor to account for {the} different branches of the
process (e.g., Refs.\ \cite{Yak2001,KYH2016}). Here we {only} need the
main dependence $Q_{\rm MU} \propto n_{p}^{1/3}$. The factor $Q_{\rm
MU\,0} \approx 1.75\times 10^{21}$~erg~cm$^{-3}$~s$^{-1}$ 
(as well as similar factors for other processes) is
calculated under the assumptions described in Ref.\ \cite{Yak2001},
with the effective masses of nucleons $m_{p}^* = 0.7 m_{ p}$ and
$m_{n}^* = 0.7 m_{n}$. The difference between the effective and bare
masses of nucleons in neutron star cores is mainly determined by {the}
many-body effects.

In the case of the DU process,
\begin{equation}
\label{eq:Qdu} Q_{\rm DU} = Q_{\rm DU\,0} \left( \frac{n_{e}}{n_0}
\right)^{1/3} T_9^6 \left( \Theta_{npe} + \Theta_{np\mu} \right),
\end{equation}
where $Q_{\rm DU\,0} \approx 1.96\times
10^{27}$~erg~cm$^{-3}$~s$^{-1}$. The factors $\Theta_{npe}$ and
$\Theta_{np\mu}$ are equal to $1$ (open the electron and muon
processes, respectively) if the Fermi momenta of the reacting particles
satisfy the corresponding triangle condition; otherwise, these
factors are zero. Because of the triangle conditions, the electron
and muon DU processes have thresholds and can operate only in the
central regions of massive neutron stars.

In the case $nn$ (of strongly superfluid protons; e.g., Ref.\
\cite{Of2015-NS}),
\begin{equation}
\label{eq:Qnn} Q_{nn} = Q_{nn \, 0} \left( \frac{n_{n}}{n_0}
\right)^{1/3} T_9^8,
\end{equation}
with $Q_{nn\,0} \approx 1.77\times 10^{19}$~erg~cm$^{-3}$~s$^{-1}$.

\subsection{Heat capacities}
\label{s:C}

Now we outline {the} specific heat capacities in neutron star cores for
the basic cases listed in Table \ref{tab:capcases} using the well-known expressions presented, for instance, in Ref.\ \cite{YakPeth2004}.

The total specific heat is
\begin{equation}
\label{eq:Cv-contrib} c_{tot} = c_{n} + c_{p} + c_{e} + c_{\mu}.
\end{equation}
For any fermion species $\alpha =n,\ldots \mu$,
\begin{equation}
\label{eq:Cva} c_{\alpha} = \frac{k_B^2}{3\hbar^3} T m_\alpha^*
p_{{F}\alpha},
\end{equation}
where $k_{ B}$ is the Boltzmann constant; $m_\alpha^*$ and
$p_{F\alpha}$ are, respectively, the effective mass and the Fermi
momentum of {the} particles $\alpha$. Note that the main contributions to
the heat capacity of nonsuperfluid cores come from {the} neutrons and
protons (e.g., Ref.\ \cite{Page1993}). Assuming again $m_{n}^* = 0.7
\,m_{n}$ and $m_{p}^* = 0.7\, m_{p}$, we obtain
\begin{equation}
\label{eq:Cvnp}
c_{N} \approx c_{0} \left( \frac{n_{N}}{n_0} \right)^{1/3} T_9,
\end{equation}
with $c_{0} = 1.12\times 10^{20}$~erg~cm$^{-3}$~K$^{-1}$.

The effective mass of ultrarelativistic, nearly ideal electrons
$m_{e}^* = p_{Fe}/c$ is determined by {the} relativistic effects. Then
\begin{equation}
\label{eq:Cve} c_{e} \approx 0.355 \,c_0\, \left( \frac{n_e}{n_0}
\right)^{2/3} T_9.
\end{equation}
Since the muons are mildly relativistic, the expression for $c_\mu$
is more complicated but the muon contribution in Eq.\
(\ref{eq:Cv-contrib}) can be roughly taken into account by an
artificial amplification of $c_e$.

\section{Grid of equations of state}
\label{sec:EOS}
\newcommand{\nliiiwr}{NL3$\omega\rho$}

\begin{figure}
\includegraphics[width=\columnwidth]{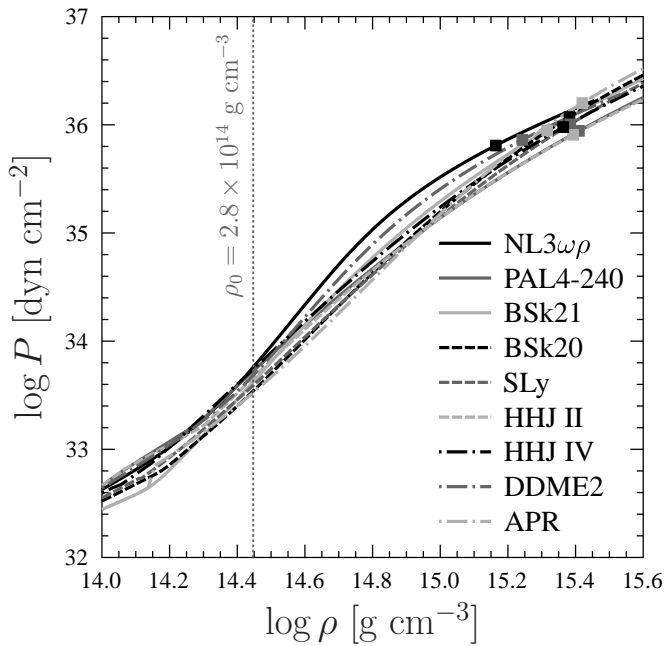}
\caption{\label{fig:P-rho} Plots of the selected
EOSs, $P=P(\rho)$, in neutron star cores. Squares mark the maximum
central densities of stable neutron stars. See text for details.}
\end{figure}
\begin{figure}
\includegraphics[width=\columnwidth]{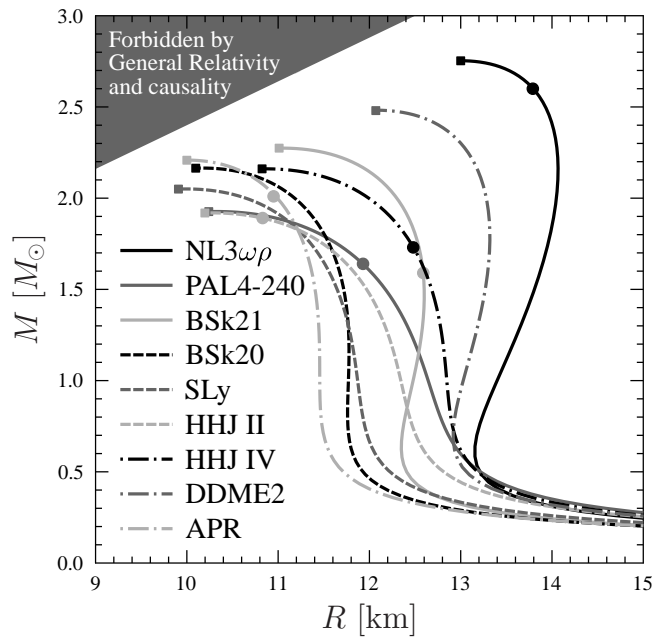}
\caption{\label{fig:M-R} $M-R$ relations for the EOSs
from Fig.\ \ref{fig:P-rho}. Filled squares mark {the} maximum masses of
stable neutron stars; filled circles mark {the} threshold masses of stars
in which the DU process is open (Table \ref{tab:EOSparams}).}
\end{figure}

\begin{table}
\begin{center}
\caption{\label{tab:EOSparams} The basic parameters of neutron stars
for the selected EOSs; $M_{\mathrm{max}}$ and $R_{\mathrm{min}}$
refer to {the} most massive stable stars; $M_{\mathrm{DU}}$ and
$R_{\mathrm{DU}}$ refer to the stars where the DU process becomes
allowed; $M_\odot$ is the mass of the Sun.}
\renewcommand{\arraystretch}{1.4}
\begin{tabular}{lcccccc}
\hline\hline EOS  &~ $M_{\mathrm{max}}$, $M_\odot$ ~&
$R_{\mathrm{min}}$, km~
& $M_{\mathrm{DU}}$, $M_\odot$~ & $R_{\mathrm{DU}}$, km \\
\hline
\nliiiwr & 2.75 & 13.00 & 2.60  & 13.79 \\
PAL4-240 & 1.93 & 10.24 & 1.64  & 11.93 \\
BSk21    & 2.27 & 11.01 & 1.59  & 12.59 \\
BSk20    & 2.16 & 10.10 &  ---  &  ---  \\
SLy     & 2.05 &  9.91 &  ---  &  ---  \\
HHJ II   & 1.92 & 10.20 & 1.89  & 10.83 \\
HHJ IV   & 2.16 & 10.82 & 1.73  & 12.48 \\
DDME2    & 2.48 & 12.07 &  ---  &  ---  \\
APR      & 2.21 & 10.00 & 2.01  & 10.95 \\
\hline\hline
\end{tabular}
\end{center}
\end{table}

\begin{figure}
\includegraphics[width=\columnwidth]{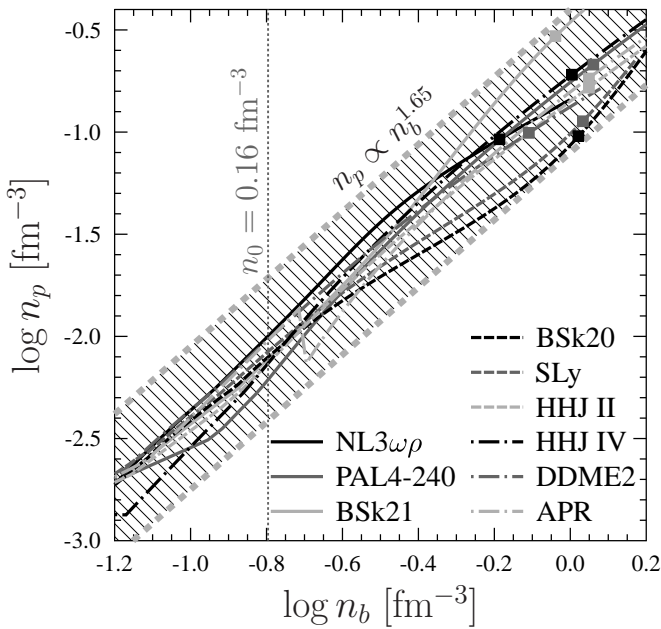}
\caption{\label{fig:np-nb} $n_{p}-n_{b}$ relations
for the selected EOSs. Squares mark the maximum $n_{b}$ in stable
neutron stars. The thick shaded strip corresponds to the power-law
models $n_{p}\propto n_{b}^{1.65}$. See text for details.}
\end{figure}

In order to calculate $L_{\nu}^\infty$ and $C_{\rm core}$, we have
selected nine EOSs of matter in neutron star cores. They are listed
in Table \ref{tab:EOSparams} and illustrated in Figs.~\ref{fig:M-R}
and \ref{fig:np-nb}. The \nliiiwr\ and DDME2 EOSs were described in
Ref.\ \cite{Fortin2016} and in references therein. The SLy EOS was
calculated in Ref.\ \cite{DH2001}. The PAL4-240 EOS was
constructed using the results of Ref.\ \cite{PA1992} but with a
different compression modulus of symmetric nuclear matter at
saturation, $K_0 = 240$~MeV (this EOS was also presented in
Appendix~D of Ref.\ \cite{HPY2007} where it was called the PAPAL
EOS). The HHJ II EOS was introduced in Ref.\ \cite{Gus2005}(where it
was called the APR~II EOS). The BSk20 and BSk21 EOSs have been
detailed and parametrized in Ref.\ \cite{BSk2013}. The HHJ IV EOS
was built in Ref.\ \cite{KKPY14}. The APR EOS was
constructed in Ref.\ \citep{APR1998}. Let us stress that the
selected EOSs are based on {\it essentially different many-body theories of dense matter}.
We have selected them to extend a class of basically different EOSs.

To construct neutron star models, one needs also an EOS in the
crust. For the SLy, BSk20 and BSk21 models, the EOSs in the crust
and the core were calculated in a unified way. The crustal EOSs
for the \nliiiwr\ and DDME2 models were described in Ref.\
\cite{Fortin2016}. The APR EOS, originally valid in the core only,
has been supplemented with the crustal part of the BSk21 EOS. For {the}
other models, {a crust EOS with a smooth composition} \cite{HPY2007}
has been used.

The selected EOSs ($P=P(\rho)$) in the neutron star cores are plotted
in Fig.\ \ref{fig:P-rho}. Near the saturation density $\rho_0$ of
the standard symmetric nuclear matter, $\rho\sim \rho_0 = 2.8\times
10^{14}$~g~cm$^{-3}$ (the dotted vertical line), they are not very
different. This is because the EOSs are usually constructed in such
a way to reproduce the properties of saturated nuclear matter which
are well studied in the laboratory.

The most important parameters of neutron stars for the selected EOSs
are listed  in Table~\ref{tab:EOSparams}. The $M(R)$ relations for {the}
neutron star models with these EOSs are plotted in
Fig.~\ref{fig:M-R}. Since we have chosen EOSs which are
sufficiently diverse and have different stiffnesses  at
$\rho\gtrsim2\rho_0$, they result in different $M(R)$ relations and
cover a large part of the $M-R$ plane. The squares in
Fig.~\ref{fig:M-R} correspond to the most massive stable neutron
stars. The selected EOSs are consistent with the recent discoveries
\cite{Demorest2010,Antoniadis2013,Fonseca} of two massive ($M \approx
2\,{\rm M}_\odot$) neutron stars. Circles mark the configurations where
the DU process becomes allowed in the neutron star center. Only
three EOSs from Table~\ref{tab:EOSparams} prohibit the DU process
for all stable neutron stars.

Figure~\ref{fig:np-nb} illustrates another important property of the
selected EOSs: the relation between the proton $n_{ p}$ and {the} total
baryon $n_{b}=n_n+n_p$ number densities. The {spread} of the curves for
the different EOSs is the thinnest at $n_{ b} \sim n_0$ (the dotted
vertical line). It is a consequence of the calibration of the EOSs
to the standard nuclear theory (see above). The straight thick
shaded strip corresponds to the relations $n_{p} \propto n_{b}^{
1.65}$. {In this way we improve the formula $n_p(n_b)$ with the
$n_{p}\propto n_{b}^2$ dependence characteristic of the
free Fermi gas model with nonrelativistic nucleons}
(e.g., Refs.\ \cite{FM1979,ST1983}). {A} smaller
power-law index effectively accounts for nucleon interactions.
According to Fig.~\ref{fig:np-nb}, this simple approximation is
qualitatively accurate; it appears sufficient for our analysis.

Note that we do not intend to accurately fit the EOSs or number
densities of {the} different particles. Our aim is to suggest some simple
scaling expressions for these quantities and use them to
describe the integral quantities,  $C_{\rm core}$ and
$L_\nu^\infty$. One can treat these scaling expressions as purely
phenomenological but we prefer to introduce them on physical
grounds. We will see that the integration over the core absorbs the
inaccuracy of {the} scaling expressions and helps {to} accurately describe
$C_{\rm core}$ and $L_\nu^\infty$ as functions of $M$ and $R$ for
the entire collection of EOSs.

\section{Calculation of the neutrino luminosity and heat capacity}
\label{sec:LnuC}

\subsection{Basic remarks}
\label{sec:basicAppr}

We have numerically calculated $L_\nu^\infty(\Tg)$ and $C_{\rm
core}(\Tg)$ from Eqs.\ (\ref{eq:L-start}) and (\ref{eq:C-start}) for
the three basic models of neutrino emission (Table
\ref{tab:nucases}) and {the} five basic models of heat capacity (Table
\ref{tab:capcases}). We have used a dense grid of neutron star
models with different $M$ for all nine EOSs from our collection
(Sec.\ \ref{sec:EOS}, Table \ref{tab:EOSparams}). In {the} calculations,
we have used accurate expressions for $Q(\rho,T)$ and $c_{\rm
core}(\rho,T)$ from Ref.~\cite{Yak2011} which are also employed in
our standard cooling code \cite{Gned2001} (taking $m_N^*=0.7 m_N$,
{to be specific}).

{Our calculation of $L_{\rm DU}^\infty(\Tg)$ deserves a comment}. As
discussed above, the DU process opens only at high densities in the
inner cores of massive stars. To simplify our analysis, we have used
$Q = Q_{\rm DU}$ throughout the entire neutron star core (to avoid
complications with the introduction of the DU threshold). This
simplification is qualitatively justified because, typically,
$Q_{\rm DU}\sim 10^6 Q_{\rm MU}$ (e.g., Ref.\ \cite{Yak2001}), and
even a small central kernel with the allowed DU process makes
$L_{\rm DU}^\infty(\Tg)$ much larger than $L_{\rm MU}^\infty(\Tg)$.
However, it somewhat overestimates $L_{\rm DU}^\infty(\Tg)$ and
{only} gives its firm upper limit. With this simplification, in all
our basic cases $L_\alpha^\infty(\Tg)$ and $C_\alpha(\Tg)$ have {a}
predetermined $\Tg$ dependence (Sec.\ \ref{s:2}), so it is
enough to choose one value of $\Tg$ and calculate
$L_\alpha^\infty(\Tg)$ and $C_\alpha(\Tg)$ for the different masses
$M$ and EOSs. The selected grid of masses was $M =
1.0\,M_{\odot},\, 1.1\,{M}_{\odot}, \ldots M_{\rm max}$. As
mentioned above, while calculating $L_{\rm DU}^\infty$ we have
extended $Q_{\rm DU}$ over the entire core, but in this case we have
not used stellar models with $M<1.5{M}_{\odot}$ because $M_{\rm DU}>
1.5~{M}_{\odot}$ for all our models (Table~\ref{tab:EOSparams}).

\subsection{Analytic approximations for $L_\nu^\infty$ and $C_{\rm core}$}
\label{sec:Integr-eval}

{The} exact analytic integration in Eqs.\ (\ref{eq:L-start}) and
(\ref{eq:C-start}) is not possible. Instead, let us derive {some}
approximate expressions for $L_\nu^\infty(\Tg)$ and $C_{\rm
core}(\Tg)$ and calibrate them using numerical results.

The first step is to assume that the main contribution to the baryon
number density $n_b = n_n + n_p$ is provided by {the} neutrons. Using a simple model
$n_{p} \propto n_{b}^{1.65}$, we get
\begin{equation}
\label{eq:np-nb} n_{n} \approx n_{b}, \quad n_{p} \approx n_{e}
\approx a n_0 \left( \frac{n_{b}}{n_0} \right)^{1.65}.
\end{equation}
Here $a$ is a dimensionless constant which we treat as a value
averaged over all selected EOSs. The approximation $n_{e} \approx
n_{p}$ can be significantly violated at very high densities at which
$n_\mu \sim n_\text{e}$ (in central regions of massive neutron stars).
Their contributions to the integrated neutrino luminosities and heat
capacities can be approximately described by an artificial
amplification of the electron contributions. For
$L_{\text{DU}}^\infty(\Tg)$, the contributions of the muon and
electron DU processes are just equal.

According to Table \ref{tab:nucases}, we study the three cases
($\alpha=nn$, MU and DU) of $L_\alpha^\infty(\Tg)$ in Eq.\
(\ref{eq:L-start}). To find an approximate expression for
$L_{nn}^\infty(\Tg)$, we take $Q=Q_{nn}$ from Eq.\ (\ref{eq:Qnn}).
In the second case ($\alpha=$MU) we employ $Q_{\rm MU}$ from Eq.\
(\ref{eq:Qmu}) with  $\Omega = \const$. In the  case $\alpha=$ DU we
use Eq.\ (\ref{eq:Qdu}) but replace the sum of $\Theta$ functions by
a factor of $2$.  Since typical densities at which the DU processes
operate are so high that the muons appear, this simplification is
reasonable.

Then $L_\nu^\infty(\Tg)$ can be evaluated with the midpoint method,
taking the integrand at some fixed value of $r=r^*$ between $0$ and
$R_\text{core}$,
\begin{equation}
\label{eq:L-midpoint} L_\nu^\infty = Q_0 R^3 \Tg_9^n \times a'
\left( \frac{R_\text{core}}{R} \right)^3 \frac{\left( n_{b*}/n_0
\right)^{k/3} \exp[(2-n)\Phi_*]}{\sqrt{1-2Gm_*/(r_* c^2)}}.
\end{equation}
Here $k=1$, $n=8$ and $Q_0 = Q_{ nn\,0}$ for $\alpha= nn$; $k=1.65$,
$n=8$ and $Q_0 = Q_{\rm MU\,0}$ for $\alpha=$MU;  $k=1.65$, $n=6$ and
$Q_0 = Q_{\rm DU\,0}$ for $\alpha=$DU. In Eq.\ (\ref{eq:L-midpoint})
we have introduced a dimensionless constant $a'$ to absorb the
inaccuracy of $L_\nu^\infty$ due to our approximations of $n_{
p}(n_{b})$ and $\Omega$ in the DU and MU cases; in the $nn$ case,
$a' = 1$. The midpoint values $n_{b*}$, $\Phi_*$ and $m_*$ are taken
at a spherical shell with $r=r_*$.

The five basic cases of heat capacity (Table \ref{tab:capcases}) can
be presented in a similar way,
\begin{equation}
\label{eq:C-midpoint} C_\text{core} = c_0 R^3 \Tg_9 \times b' \left(
\frac{R_\text{core}}{R} \right)^3 \frac{\left( n_{{b}*}/n_0
\right)^{k/3} \exp(-\Phi_*)}{\sqrt{1-2Gm_*/(r_* c^2)}}.
\end{equation}
According to Eq.\ (\ref{eq:Cvnp}), in the case of $C_n$ we have
$k=1$ and $b'=1$. In the case of $C_p$ from Eq.\ (\ref{eq:np-nb}) we
employ $k=1.65$. As for the case of $C_\ell$, we assume that the
main contribution comes from {the} electrons, Eq.\ (\ref{eq:Cve});
using Eq.\ (\ref{eq:np-nb}) we set $k=2\times 1.65 = 3.30$. The
constants $b'$ in the two latter cases are thought to absorb the
inaccuracies of these analytic approaches. The heat capacities
$C_{tot}$ and $C_{n\ell}$ are thought to be mainly determined by the
neutrons. Then we set $k=1$ for both cases and assume that tuning
$b'$ will make the approximations sufficiently accurate.

The next step is to consider a polytropic EOS model, $P = K
\rho^\gamma$, which is a primitive but useful approximation. Since
all particles in the core are strongly degenerate,
{\begin{equation} \label{eq:P-rhonb} 
{P} = c^2 n_b^2 \frac{{\rm d} (\rho/n_{b})}{{\rm d} n_{b}}.
\end{equation}}
Then, using the polytropic relation and assuming the boundary
{condition ${\rm d}\rho/{\rm d} n_{b} = m_0$ at the neutron star
surface we get, for catalyzed neutron star matter,}
\begin{equation}
\label{eq:nb-rho} n_{b} = \frac{\rho}{m_0} \left( 1 +
\frac{K}{c^2}\rho^{\gamma-1} \right)^{-1/(\gamma-1)}.
\end{equation}
Here $m_0$ is the {mass} per baryon in the $^{56}$Fe {atom}. For any
neutron star, we can find $\rho_*$, $K$ and $\gamma$ to evaluate
$n_{b*}$ needed in Eqs.\ (\ref{eq:L-midpoint}) and
(\ref{eq:C-midpoint}). Note that $\gamma$ is {specific to} a given
star; it can vary with growing $M$.

Now we are ready to evaluate the relation between $\Phi_*$ and
$\rho_*$. Using Eq.\ (\ref{eq:nb-rho}) one can solve Eq.\
(\ref{eq:dPhi}) and obtain
\begin{equation}
\label{eq:Phi-rhoPnb} \Phi = \frac{1}{2}\ln(1-x_{g}) - \ln \left(
\frac{P+\rho c^2}{m_0 c^2 n_{b}} \right).
\end{equation}
Here we have used the boundary value $\Phi_s = \frac{1}{2}\, \ln (1-x_{g})$ at
the stellar surface, with $x_{g} = 2GM/(Rc^2)$. The polytropic
approach yields
\begin{equation}
\label{eq:PhiStar} \exp \Phi_* = \sqrt{1-x_{g}} \left( 1 +
\frac{K}{c^2}\,\rho_*^{\gamma-1} \right)^{-\gamma/(\gamma-1)}.
\end{equation}

The quantity $\xi(M,R)=R/R_\text{core}$ can be taken from Refs\
\cite{HZ2011,ZH2016} as
\begin{equation}
\label{eq:RcoreR}
\xi(M,R)\equiv\frac{R}{R_\text{core}} = 1/x_g -
\exp(-2\chi_{cc})(1/x_g-1),
\end{equation}
where $\chi_{cc}=\int_0^{P_{cc}} {\rm d}P/(P+\rho c^2)$ is an
integral over the crust and $P_{cc}$ is the pressure at the
core/crust interface. For catalyzed matter, one has
$\exp{(\chi_{cc})}=\mu_{cc}/m_0c^2$, where $\mu$ is {the} baryon
chemical potential. The value $\chi_{cc}$ slightly varies from one
EOS to another. Here we adopt $\chi_\text{cc}=0.03$ as a unified
value for all the EOSs of our study.

\begin{table*}
\begin{center}
\caption{\label{tab:npkabcgamma} Parameters of the approximations
(\ref{eq:Jdef}) and (\ref{eq:LC-final}) for the three basic models
of neutrino luminosity $L_{\nu}^\infty$ and {the} five basic models
of
 heat capacity $C_{\rm core}$.}
\renewcommand{\arraystretch}{1.4}
\begin{tabular}{ccccccccccccc}
\hline\hline $L_\nu^\infty$ or $C_{\rm core}$ & Case & $Q_0^\dag$ or
$c_0^\ddag$ & ~$n$ & ~$k$ & ~$p$ & ~$a_1$ & ~$a_2$ & ~~$a_3$~~ &
~~$a_4$~~ & ~~$a_5$~~ & ~rms~ & max error \\
\hline
                 & $nn$ & $1.77\times 10^{19}$ & 8 & 1 & 6 & 3.54 & 0.0125 & 2.73 &  4.33  & 0.509 & 0.05 & 0.17 \\
$L_{\nu}^\infty$ & MU & $1.75\times 10^{21}$ & 8 & 1.65 & 6 & 2.05 & 0.0125 & 2.58 & 4.40 & 0.480 & 0.15 & 0.42 \\
                 & DU & $1.96\times 10^{27}$ & 6 & 1.65 & 4 & 1.80 & 0.0070 & 2.62 & 4.80 & 0.501 & 0.08 & 0.20 \\
\hline
                     & $n$ & $1.12\times 10^{20}$ & 1 & 1    & 1 & 2.86 & 0.0119 & 2.49 & 3.68 & 0.408 & 0.0084 & 0.025 \\
                     & $p$ & $1.12\times 10^{20}$ & 1 & 1.65 & 1 & 0.781 & 0.0069 & 2.70 & 5.75 & 0.657 & 0.062 & 0.17 \\
$C_{\rm core}$ & $\ell$ & $1.12\times 10^{20}$ & 1 & 3.30 & 1 & 0.0823 & 0.0033 & 2.60 & 5.00 & 0.800 & 0.14 & 0.31 \\
           & $tot$ & $1.12\times 10^{20}$ & 1 & 1    & 1 & 4.17 & 0.0130 & 2.59 & 3.50 & 0.800 & 0.023 & 0.075 \\
             & $n \ell$ & $1.12\times 10^{20}$ & 1 & 1    & 1 & 3.01 & 0.0130 & 2.59 & 3.50 & 0.799 & 0.015 & 0.047 \\
\hline\hline
\end{tabular}\\
$^\dag$erg~cm$^{-3}$~s$^{-1}$; $^\ddag$erg~cm$^{-3}$~K$^{-1}$
\end{center}
\end{table*}

To approximate $L_\nu^\infty(\Tg)$ and $C_\text{core}(\Tg)$ we
should substitute Eqs.\ (\ref{eq:nb-rho}), (\ref{eq:PhiStar}) and
(\ref{eq:RcoreR}) into Eqs.\ (\ref{eq:L-midpoint}) and
(\ref{eq:C-midpoint}) and fix $a'$, $b'$, $\rho_*$, $m_*$, $r_*$,
$\gamma$ and $K$. Let us introduce $x_\rho = M/(\rho_0 R^3)$ and
assume that $\rho_* \sim M/R_\text{core}^3$. Thus the approximations
are
\begin{subequations}
\label{eq:recipe}
\begin{eqnarray}
\label{eq:recipe-a1}
\left\{ \begin{array}{c}
a' \\
b'
\end{array} \right\} \left( \frac{\rho_*}{m_0 n_0} \right)^{k/3} & \to & a_1 \xi^k x_\rho^{k/3}, \\
\label{eq:recipe-a2}
K \rho_*^{\gamma-1}/c^2 & \to & \left( a_2 x_\rho \xi^3 \right)^{\gamma-1}, \\
\label{eq:recipe-a34}
\gamma & \to & \frac{a_3}{1 + a_4\xi\sqrt{{x_{g}^5/x_{\rho}}}}, \\
\label{eq:recipe-a5} \frac{2Gm_*}{r_*c^2} & \to & a_5 x_{g},
\end{eqnarray}
\end{subequations}
where $a_1,\ldots ,a_5$ are the fit parameters to be optimized. They
are expected to be different for {the} different quantities (for
$L_\nu^\infty$ due to different neutrino processes, and for
$C_\text{core}$ due to different particle fractions).

Note that the expression for $\gamma$ is quite arbitrary to
account for the fact that more massive and, consequently, denser
stars should contain softer matter. In Eq.\ (\ref{eq:recipe-a34}),
$\gamma$ depends actually on $\xi\sqrt{x_{g}^5/x_\rho} \propto
M^2/R_\text{core}$, which ensures its reasonable dependence.

It is convenient to introduce
\begin{equation}
\label{eq:Jdef}
  J_{kp}(M,R) =  a_1 \xi^{k-3} \frac{x_\rho^{k/3}
  \left[ 1 + \left( a_2 x_\rho \xi^3 \right)^{\gamma-1}\right]^{\frac{p\gamma - k/3}{\gamma-1}}}
  {\left( 1 - x_{g} \right)^{p/2} \sqrt{1 - a_5 x_{g} }}
\end{equation}
with $\xi$ given by Eq.\ (\ref{eq:RcoreR}) and $\gamma$ by Eq.\
(\ref{eq:recipe-a34}). Finally, the approximations take the forms
\begin{equation}
\label{eq:LC-final}
\left\{ \begin{array}{c}
L_{\nu}^{\infty}(\Tg) \\
C_{\rm core}(\Tg)
\end{array}\right\}
=  \left\{ \begin{array}{c}
Q_0 \\
c_{0}
\end{array}\right\}
R^3 \Tg_9^n J_{kp}(M,R).
\end{equation}
The values $n$, $p$ and $k$ are taken from Eqs.\
(\ref{eq:L-midpoint}) and (\ref{eq:C-midpoint}) and listed in
Table~\ref{tab:npkabcgamma}. The dimensionless parameters
$a_1,\ldots, a_5$ in Eq.\ (\ref{eq:Jdef}) will be obtained by the
calibration to numerical calculations.

\subsection{Calibration to numerical calculations}
\label{sec:Integr-calibr}

\begin{figure*}
\includegraphics[width=\textwidth]{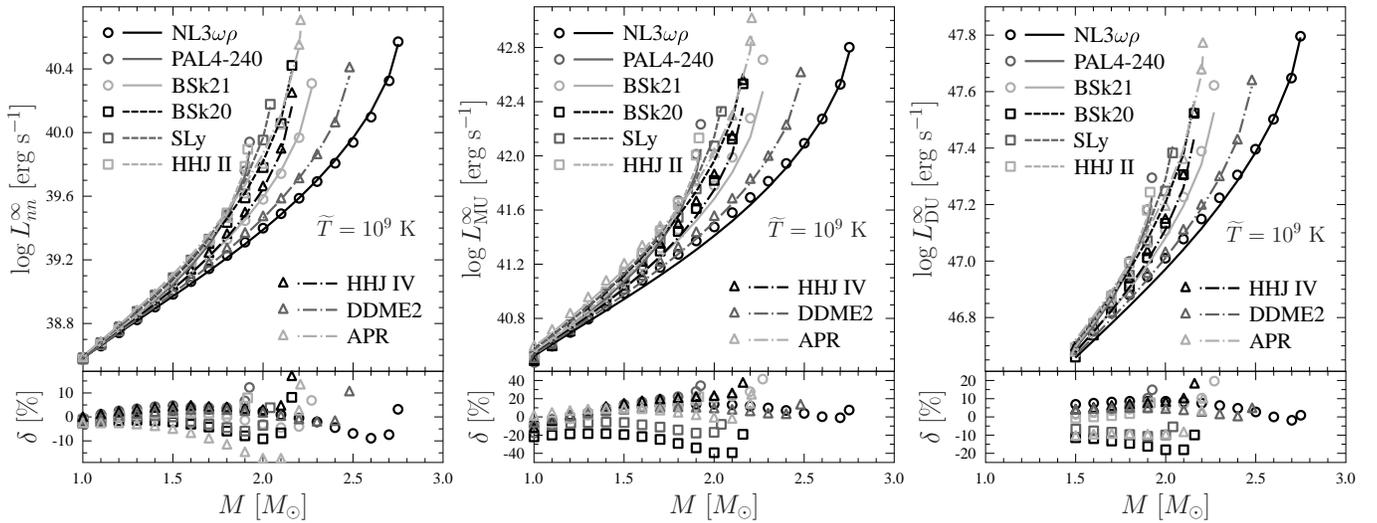}
\caption{\label{fig:M-lgL} $L_\nu^\infty-M$ relations
for the nine selected EOSs at $\Tg = 10^9$~K. Lines show the
approximation (\ref{eq:LC-final}); squares, circles and triangles
show numerical calculations. For $L_{\rm DU}^\infty$, the DU
process is artificially extended over the entire core, but the
calculations are performed only at $M\geq 1.5M_\odot$. The bottom
panels display the relative fit errors. See text for details. }
\end{figure*}

\begin{figure*}
\includegraphics[width=\textwidth]{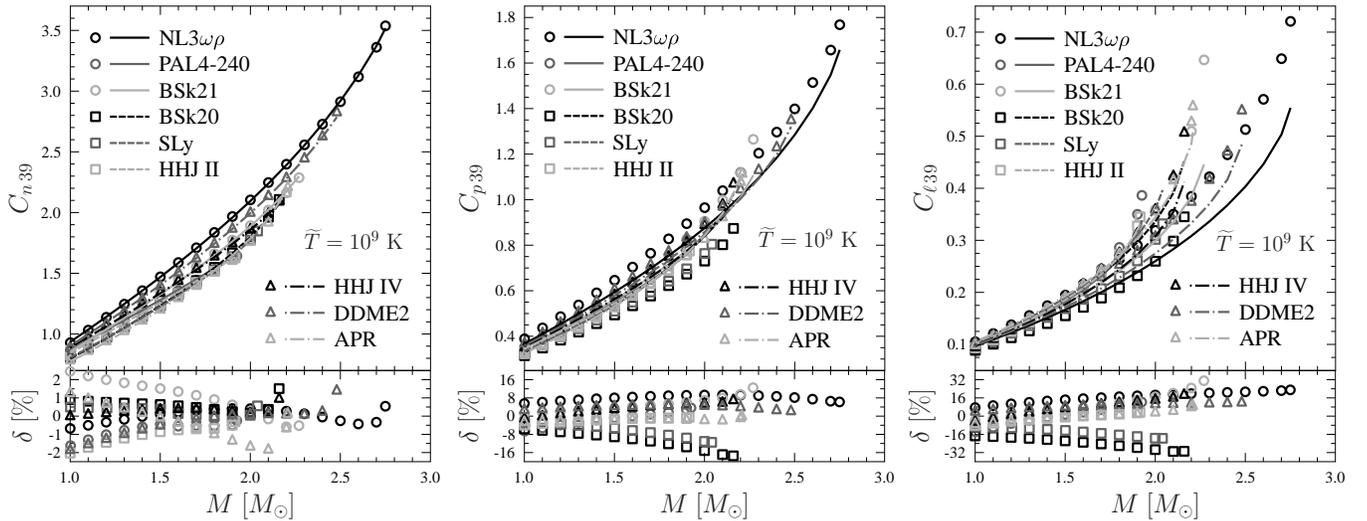}
\caption{\label{fig:M-C39npl} $C_{\rm core}-M$
relations for the nine selected EOSs at $\Tg = 10^9$~K, with
$C_{\alpha\,39} = C_\alpha/(10^{39}\,\text{erg K$^{-1}$})$ for
$\alpha = n, p$ and  $\ell$. Lines show the approximation
(\ref{eq:LC-final}); squares, circles and triangles show
numerical calculations. The bottom panels display the relative fit
errors. See text for details.}
\end{figure*}

\begin{figure}
\includegraphics[width=\columnwidth]{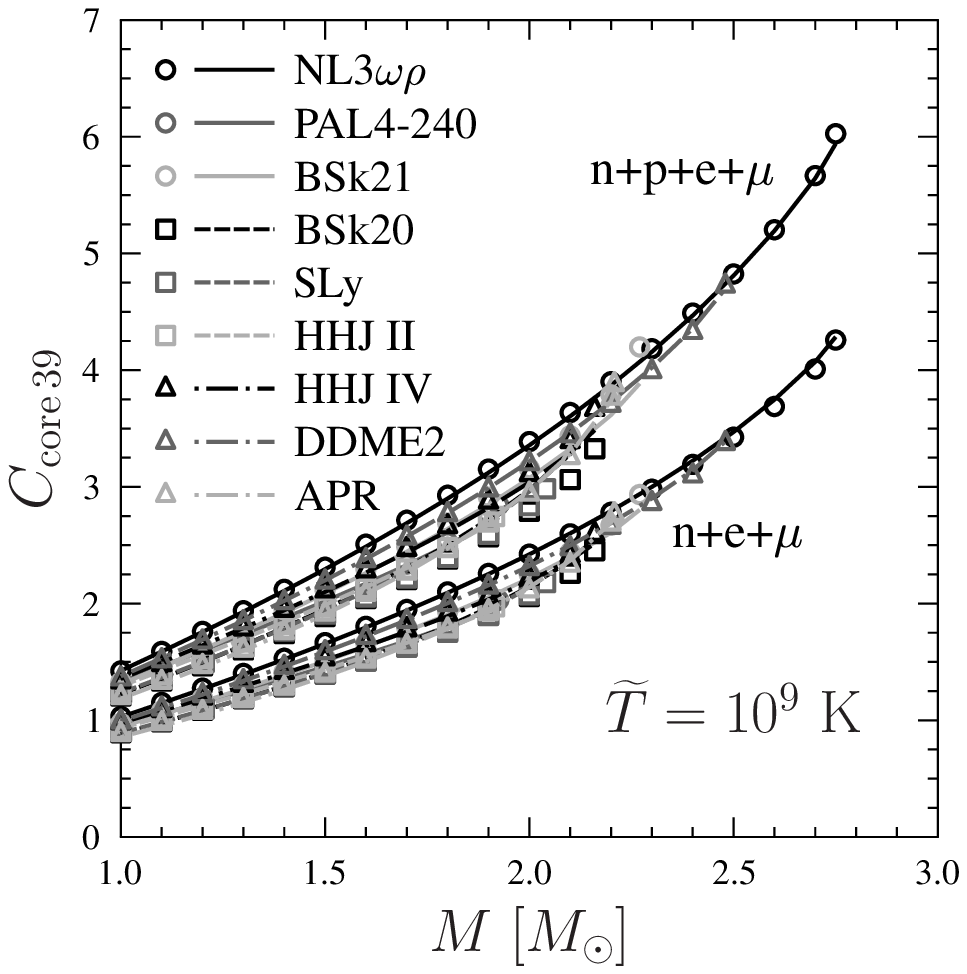}
\caption{\label{fig:M-C39normSF} $C_{\rm core}$ vs.\
$M$ for the nine selected EOSs at $\Tg = 10^9$~K;
$C_{\text{core}\,39} = C_\text{core}/(10^{39}\,\text{erg
K$^{-1}$})$. The legend is the same as in Fig.~\ref{fig:M-C39npl}.
The``n+p+e+$\mu$'' = ``{\it tot}'' curves correspond
to the fully non-superfluid core.
The ``n+e+$\mu$'' = ``$n\ell$'' curve is for
the core with strongly superfluid protons
but normal neutrons. See text for details.}
\end{figure}

Our numerical results (Sec.\ \ref{sec:basicAppr}) are shown by
different symbols in Figs.~\ref{fig:M-lgL}--\ref{fig:M-C39normSF}.
We have obtained 123 values of $L_\alpha^\infty$ (for each
$\alpha=nn$ and MU) as well as 123 values of $C_\alpha$ (for each
$\alpha=n,p,\ell,tot,nl$). For $L_{\rm DU}^\infty$, we have excluded
45 values with $M<1.5~M_\odot$. The trial functions
$L_{\nu}^{\infty}(M,R)$ and $C_{\rm core}(M,R)$ (Eqs.\
(\ref{eq:Jdef}) and (\ref{eq:LC-final})) have been calibrated to
these data sets. The target function to be minimized has been the
relative root mean square (rms) error. We present also the maximum
relative fit errors over thesame data sets.
The parameters $Q_0$, $c_0$, $n$, $p$ and $k$ have
been taken fixed from the consideration presented above. We have
varied $a_1,\ldots,a_5$ to minimize the rms error. We present the values of
the these parameters with minimum number of digits which
do not change the rms error taken with two significant
digital numbers. The optimized values of
$a_1,\ldots ,a_5$ as well as the corresponding fit errors are listed
in Table~\ref{tab:npkabcgamma}.
Figures~\ref{fig:M-lgL}--\ref{fig:M-C39normSF} compare the fits with
numerical calculations.

Let us discuss the approximations of $L_{\nu}^\infty$. They are
{the} most precise for the nn-bremsstrahlung; the rms error appears
to be the lowest here because $Q_{ nn}$ is independent of the
fractions of charged particles in dense matter. The largest errors
occur for the MU case due to a strong dependence of $Q_{\rm MU}$ on
the fractions of charged particles through the factor $\Omega$. The
approximation of $L_{\rm DU}^\infty$ is more accurate than that of
$L_{ \rm MU}^\infty$ because $Q_{\rm DU}$ depends on $n_{e}$ in a
relatively simple way.

The importance of {the} charged particle fractions can be
demonstrated by {the} instructive examples of the BSk20 and HHJ~IV
EOSs. In Figs.~\ref{fig:M-R}--\ref{fig:M-lgL} the
corresponding curves are plotted by the blue short-dashed
(BSk20) and black dot-dashed (HHJ~IV) lines. The numerical
data in Fig.~\ref{fig:M-lgL} are displayed
by the black squares (BSk20) and black triangles (HHJ~IV). According to
Fig.~\ref{fig:M-R}, these EOSs result in very close maximum masses,
but the stars with the BSk20 EOS are more compact, i.e. they have smaller
radii than the HHJ~IV stars of the same $M$. Roughly speaking, the
$M-R$ relations for these EOSs differ by a shift along the $R$ axis.
This means that a BSk20 star is denser than an HHJ~IV star, and,
therefore, has larger $L_\nu^\infty$. This is really true for $L_{
nn}^\infty$ (Fig.~\ref{fig:M-lgL}): the
squares (for the BSk20 EOS) lie higher than the
triangles (for the HHJ~IV EOS). This feature is well reproduced by
the corresponding black dashed and dot-dashed lines,
which show the approximations
(\ref{eq:LC-final}) for these EOSs. In contrast, the MU and DU
luminosities are sensitive to the $n_{p} (n_{b})$ relations.
According to Fig.~\ref{fig:np-nb}, the values of $n_{p}$ for the
HHJ~IV EOS are noticeably higher than for the BSk20 EOS. The
opposite effects of the two factors, the greater compactness of the
BSk20 stars and the larger $n_{p}$ for the HHJ~IV stars, lead to
their compensation. Accordingly, the DU as well as the MU neutrino
luminosities for these EOSs appear to be close enough (the
corresponding triangles and squares in the middle and left panels of
Fig.~\ref{fig:M-lgL} overlap). Because the approximation
(\ref{eq:LC-final}) is derived using a not very accurate description
of {the} proton, electron and muon fractions, it cannot reproduce
this effect exactly; an approximate expression gives $L_{\rm
DU}^\infty$ and $L_{\rm MU}^\infty$ higher than {the} numerical
values for the BSk20 EOS and lower than for the HHJ~IV EOS.
Moreover, the MU and DU luminosities of the BSk20 and SLy stars are
systematically overestimated by the approximation
(\ref{eq:LC-final}) as these EOSs have essentially smaller charged
particle fractions than the other EOSs.

Now let us outline the approximations of the heat capacity
(Figs.~\ref{fig:M-C39npl} and \ref{fig:M-C39normSF}). The neutron
contribution (the left panel of Fig.~\ref{fig:M-C39npl}) is
accurately reproduced by the approximation (\ref{eq:LC-final}). It
is precise enough to distinguish {between} very close $C_n-M$
relations for different EOSs. On the contrary, the approximations
hardly resolve {the} different curves for $C_p$ and $C_\ell$. The
fit errors are about seven times larger for $C_p$ and 12--15 times
larger for $C_\ell$ since the details of the $n_p(n_{b})$ and
$n_{e}(n_{b})$ relations cannot be well reproduced by the functions
which depend on $M$ and $R$ only. Similarly to $L_\text{MU}^\infty$
and $L_\text{DU}^\infty$, {the} numerical values of $C_p$ and
$C_\ell$ for the BSk20 and SLy stars are systematically smaller than
the fitted values. Nevertheless, since $C_n$ dominates over {the}
other contributions, the approximation (\ref{eq:LC-final}) almost
precisely reproduces $C_{tot}$ and $C_n$
(Fig.~\ref{fig:M-C39normSF}). The difference between the $a_1$
values shows that switching off the proton contribution reduces
$C_\text{core}$ by about 25\%, in agreement with the results of
Ref.\ \cite{Page1993}. Note that {the} sum of the fits
$C_n+C_p+C_\ell$ gives 1--2\% larger errors than $C_{tot}$, with the
parameters from the last two lines in Table~\ref{tab:npkabcgamma}.

Let us mention several common features of our approximations. First,
the index $\gamma$ given by Eq.\ (\ref{eq:recipe-a34}) with the
values of $a_3$ and $a_4$ from Table~\ref{tab:npkabcgamma} ranges
from 2.3--2.5 for low-mass stars to 1.7--1.9 for high-mass stars.
This seems realistic for the considered set of EOSs. Second, the fit
errors increase with growing $M$ (except for the almost precise
approximation of the neutron heat capacity), because the higher the
density, the stronger the difference between the EOS models.

Our grid of selected EOSs is wide but nevertheless restricted.
For instance, $M_{\rm DU}>1.5 \, M_\odot$ 
for all of them (Table \ref{tab:EOSparams}), which 
seems reasonable (e.g., Ref.\ \cite{Bez2015a} and references therein) 
but is not strictly proven. To check ``universality'' of our
fits we have taken the NL3 EOS \cite{Fortin2016} (with
$M_{\rm DU}=0.84 M_\odot$ and $M_{\rm max}=2.77 M_\odot$).
Some fit errors appear to be somewhat higher while others
somewhat lower than for the selected EOSs;  
nevertheless they seem acceptable. For instance, the maximum relative
errors in the $nn$, MU and DU cases for stars with the NL3 EOS 
appear to be 0.07,  0.39 and 0.22, respectively, while for
the selected EOSs we had 0.17, 0.42, and 0.29 (Tables 
\ref{tab:EOSparams} and \ref{tab:npkabcgamma}).
For the heat capacities $n$, $\ell$, $p$, $tot$, $n\ell$ we now
obtain the maximum errors  0.09, 0.22, 0.47, 0.09, and 0.07 versus
0.025, 0.17, 0.31, 0.075, 0.047 for the selected EOSs.

\section{Importance of redshift in neutron star interior}
\label{sec:Redshift}

\begin{figure}
\includegraphics[width=\columnwidth]{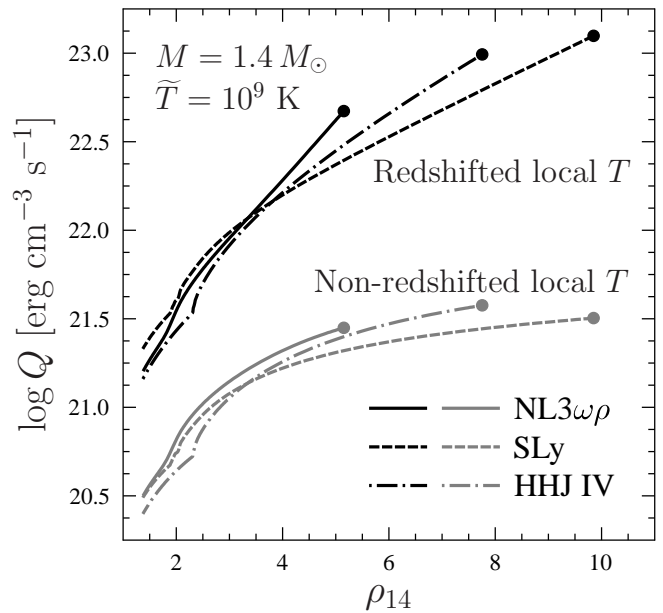}
\caption{\label{fig:Q(rho)} Logarithm of the neutrino emissivity
$Q=Q_{\rm MU}$ (standard neutrino candle) versus
$\rho_{14}=\rho/(10^{14}~{\rm g}~{\rm cm}^{-3})$ in the core of a
$1.4\,M_\odot$ neutron star with three different EOS models. The
core is isothermal, with $\Tg=10^9$~K. The upper (black) lines are
calculated using {the} correctly redshifted local temperature $T$ of
the matter. The lower (gray) lines are calculated neglecting the
gravitational redshift, that is with $T=\Tg$. A jump of $Q$ at lower
$\rho$ is due to the appearance of muons. Filled circles mark $Q$
and $\rho$ in the center of the star.}
\end{figure}

\begin{figure}
\includegraphics[width=\columnwidth]{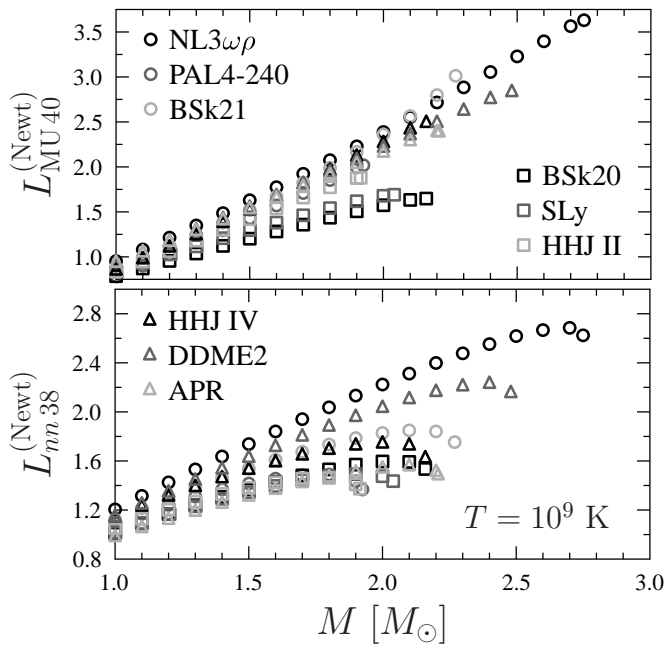}
\caption{\label{fig:Newt} $L_\nu^\text{(Newt)}$ vs.\ $M$ for $nn$-bremsstrahlung
(the bottom panel, in $10^{38}$~erg~s$^{-1}$) and MU (the top panel,
in $10^{40}$~erg~s$^{-1}$) processes for the nine selected EOS models at $\Tg = 10^9$~K.}
\end{figure}

While calculating $L_\nu^\infty$ and $C_\text{core}$, it is
important to include the proper temperature and emissivity redshifts
in a neutron star core.

For instance, Fig.\ \ref{fig:Q(rho)} shows the neutrino emissivity
$Q=Q_{\rm MU}(\rho)$ of the standard neutrino candle (due to the MU
process) in {the} nonsuperfluid isothermal core of a neutron star.
The star is assumed to have the BSk21 EOS and the mass
$M=1.4\,M_\odot$. The isothermal core temperature is taken to be
$10^9$ K (just for example). The black line is calculated correctly,
with $\Tg=10^9$ K and the local temperature of the matter $T=\Tg
\,\exp(-\Phi)$ which is higher, than the isothermal temperature
$\Tg$. The gray line is calculated neglecting the gravitational
redshift, using (erroneously and intentionally) a constant local
temperature $T=10^9$ K throughout the core. A kick of the neutrino
{emissivity} at lower $\rho$ is caused by the appearance of muons
and the associated muonic MU process (e.g., Ref.\ \cite{Yak2001}).
One can see that the General Relativistic redshift is very
significant for correctly calculating the neutrino
luminosity $L_\nu^\infty$. It greatly increases the local
temperature and the neutrino emissivity in the core. This effect
becomes stronger near the center of the star (where the redshift is
larger). Therefore, the redshift effect strongly intensifies the
integrated neutrino luminosity $L_\nu^\infty$ and increases the
contribution of the central part of the core to $L_\nu^\infty$.

In order to demonstrate this we have calculated the integrals
(\ref{eq:L-start}) and (\ref{eq:C-start}) ignoring all the factors
$\exp \Phi$. Specifically, this means that we use the ``Newtonian''
thermodynamic equilibrium, $T = const$. We will call such artificial
quantities ``Newtonian'', $L_\nu^\text{(Newt)}$ and
$C_\text{core}^\text{(Newt)}$, while the true quantities
(\ref{eq:L-start}) and (\ref{eq:C-start}) will be called
``relativistic''.

Figure~\ref{fig:Newt} plots the $nn$ and MU ``Newtonian''
luminosities. Comparing them with the left and middle panels of
Fig.~\ref{fig:M-lgL} we see that $L_\nu^\infty$ and
$L_\nu^\text{(Newt)}$ differ as functions of $M$. If $M$ varies from
$1\,M_\odot$ to $\sim 3\,M_\odot$, the ``Newtonian'' ones change only
by a factor of two or three, while the ``relativistic'' luminosities
change by two orders of magnitude. Moreover, $L_{nn}^\text{(Newt)}$
becomes nonmonotonic near the maximum masses due to a dramatic
decrease of {the} stellar radius. This phenomenon vanishes for the
``relativistic'' $nn$ luminosity because the total redshift
factor $\exp(-6\Phi)$ in
the integrand of Eq.\ (\ref{eq:L-start}) becomes very large ($\sim
10 - 100$) for high-mass stars. Another feature is an inverted
ordering of {the} ``Newtonian'' $L_{nn}^\infty$ with respect to the
``relativistic'' ones for a fixed value of $M$. Note that this
ordering completely breaks for $L_{\rm MU}^\text{(Newt)}$ because it
depends on the fraction of charged particles. It is rather close to
the ordering of {the} proton number densities at a fixed $n_{b}$ in
Fig.~\ref{fig:np-nb}. Nevertheless, the ordering of the
``relativistic'' MU luminosities in Fig.~\ref{fig:M-lgL} does not
correlate with the ordering of $n_{p}-n_{b}$ curves in
Fig.~\ref{fig:np-nb}.

Note that the neutron contribution to $C_\text{core}^\text{(Newt)}$
is given by the same integral as $L_{ nn}^\text{(Newt)}$ but with
$Q_{ 0\, nn}$ from Eq.\ (\ref{eq:Qnn}) replaced by the expression
for $c_0$ from Eq.\ (\ref{eq:Cvnp}). Thus the bottom panel of
Fig.~\ref{fig:Newt} looks like $C_{n}^\text{(Newt)}$. Its difference
from the `relativistic' quantity (the left panel of
Fig.~\ref{fig:M-C39npl}) is not so dramatic as for neutrino
luminosities, but $C_\text{core}$ given by
Eq.~(\ref{eq:C-start}) has a concave-shaped mass dependence, while
the curves of the `Newtonian' heat capacity --- mass relation is
convex.

Our analysis shows that the true ``relativistic'' luminosities
and heat capacities are dominated by
the General Relativistic redshift, as well as by the effects of
nuclear and particle physics, like the $n_{p}-n_{b}$ relation
and the emissivity dependence on particle number densities.

\newcommand{\B}{\mathcal{B}}
\newcommand{\Msun}{M$_\odot$}

\begin{table*}
\begin{center}
\caption{{\label{tab:obs}} Observational data on nine selected isolated neutron stars which are at the neutrino cooling stage.}
\small
\renewcommand{\arraystretch}{1.4}
\begin{tabular}{lcccccc}
\hline\hline
Name  &  $t$ [kyr]  &  $T_\text{\rm s}^\infty$ [MK]  &  $M$ [$M_\odot$]  &  $R$ [km]  &   $B$ [TG] & Ref. \\
\hline PSR J0205+6449 (in 3C 58) & $0.82-5.4$ & $<1.02$      & 1.4 &
12 & 3.6 & \cite{SHSM04, Shibanov_etal08}\\
PSR B0531+21 (Crab)       & $1.0$      & $<2.0$       & 1.4 & 12.14 & 3.8 & \cite{Weisskopf_etal04,   Weisskopf_etal11}\\
PSR J1119--6127           & $0.8-3.2$  & $1.02-1.48$  & 1.4 & 10    &  41 & \cite{Zavlin09} \\
RX J0822--4300 (in Pup A) & $3.6-5.2$  & $1.6-1.9$    & 1.4 & 10    & 0.033 & \cite{ZTP99, BPWP12,2009GH} \\
PSR J1357--6429           & $3.65-14.6$& $0.68-0.86$  & 1.4 & 10    &  7.8 & \cite{Zavlin07} \\
PSR B0833--45 (Vela)      & $11-25$    & $0.65-0.71$  & 1.4 & 10    &  3.4 & \cite{Pavlov_etal01} \\
PSR B1706--44             & $8.5-34$   & $0.48-0.83$  & 1.4 & 12    &  3.1 & \cite{McGowan_etal04} \\
PSR J0538+2817            & $26-34$    & $0.71-1.07$  & 1.4 & 10.5  &  0.7 & \cite{ZP04} \\
PSR B2334+61              & $20.5-82$  & $0.55-0.84$  & 1.4 & 10    &  9.9 & \cite{Zavlin09} \\
\hline\hline
\end{tabular}
\end{center}
\end{table*}

\section{Illustrative examples}
\label{sec:cooling}

\begin{figure*}
\includegraphics[width=1.0\textwidth]{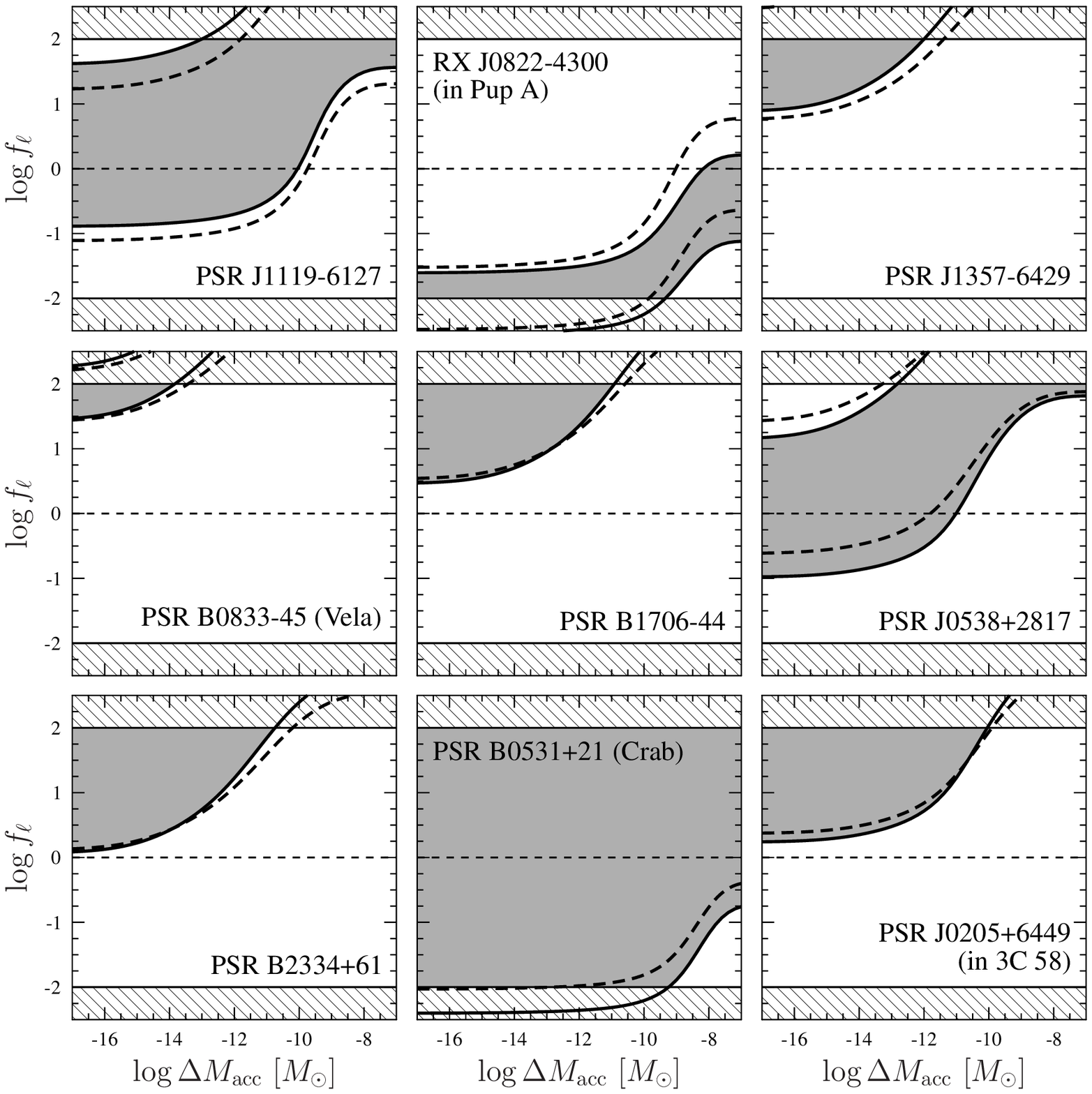}
\caption{\label{fig:AFigura} Neutrino cooling function $f_\ell$
vs mass $\Delta M_{\rm acc}$ of light accreted elements for nine
isolated neutron {stars at} the neutrino cooling stage (Table
\ref{tab:obs}). The hatched regions exclude too low ($f_\ell<0.01$)
and too high ($f_\ell>100$) neutrino emission levels which are
unreasonable from {the} theoretical point of view. The short dashed
lines refer to the standard neutrino emission level (standard
neutrino candle). The grayed regions (restricted mainly by the thick
solid lines) show {the} theoretically accessible ranges of $f_\ell$
and $\Delta M_{\rm acc}$ calculated taking into account {the}
surface magnetic fields. The thick long-dashed lines display the
same bound as the thick solid lines but neglecting the effects of
$B$ fields. See text for details.}
\end{figure*}

To illustrate our results let us outline a scheme for a possible
interpretation of {the} observations of neutron stars. We have
selected nine isolated middle-aged cooling neutron stars listed in
Table \ref{tab:obs}. They are PSR J1119--6127 (hereafter J1119), RX
J0822--4300 (in Pup~A), PSR J1357--6429 (J1357), PSR B0833--45 (Vela),
PSR B1706--44 (B1706), PSR J0538+2817 (J0538), PSR B2334+61 (B2334),
PSR B0531+21 (Crab) and PSR J0205+6449 (in 3C\ 58).

For each of these stars, the ages $t$ and the effective surface
temperatures $T_{s}^\infty$ have been estimated or constrained.
These observational data are presented in Table \ref{tab:obs}
together with the references from which the data are taken. Note
that the surface temperature of the Crab pulsar and the pulsar in
3C\ 58 is {an} upper limit only. The surface
temperatures have been inferred from
{the} observed spectra of neutron stars using either {a} blackbody
model, or hydrogen atmosphere models (magnetic or nonmagnetic
ones). In performing such analyses we have adopted certain values of
neutron star mass $M$ and radius $R$ which are also listed in the
table. Finally, the magnetic fields of these stars have been mainly
estimated from {the} standard magnetic braking measurements. The
values of the surface equatorial magnetic field $B$ were
taken from the ATNF pulsar catalogue\footnote{
www.atnf.csiro.au/people/pulsar/psrcat/}
\cite{MANCHESTER2005}.

All these neutron stars are thought to be at the neutrino cooling
stage with isothermal interiors. Their cooling is regulated by the
neutrino cooling function $L_\nu^\infty(\Tg)/C_{\rm core}(\Tg)$
which can be easily calculated from our approximations of
$L_\nu^\infty(\Tg)$ and $C_{\rm core}(\Tg)$ (Sec.\ \ref{sec:LnuC}).
Here we adopt the minimal cooling paradigm \cite{PageMinCool2004,
2004Gusakov}, according to which neutron star cores consist of
nucleons, muons and electrons (our main assumption throughout this
paper) and the DU process is not open there. Following the method
developed in our previous studies \cite{Weisskopf_etal11, Yak2011,
Kloch2015, Of2015-NS}, we can assume certain values of $t$,
$T_s^\infty$, $M$, $R$, and the mass $\Delta M_{\rm acc}$ of light
elements in the heat blanketing envelope and infer the dimensionless
cooling function of the neutron star we are studying,
\begin{equation}
   f_\ell= \frac{L_\nu^\infty/C_{\rm core}}{L_{\rm MU}^\infty/C_{tot}},
\label{eq:fl}
\end{equation}
which is the ratio of the actual cooling function to the function
for the standard neutrino candle (that is a star of the same mass,
radius and EOS but without superfluidity in the core). The magnitude
of $f_\ell$ is a fundamental parameter of the superdense matter in
neutron stars. For a standard candle, $f_\ell=1$. Its realistic
minimal value is $f_\ell \sim 0.01$. It is realized in the presence
of strong proton core superfluidity that drastically suppresses the
MU process (leaving the $nn$ bremsstrahlung to be the leading
neutrino emission mechanism in the core). Its realistic maximum
value $f_\ell \sim 100$ is realized in the presence of triplet-state
neutron pairing which is accompanied by a strong Cooper pairing
neutrino emission. Note that the method of extracting $f_\ell$ from
the observations is not very accurate because of the strong
temperature dependence of $L_\nu^\infty/C_{\rm core}$
\cite{Of2015-NS}.

Therefore, the realistic values of $f_\ell$ range from $\sim 0.01$
to $\sim 100$. Since we do not consider the DU process, in many
cooling scenarios $f_\ell$ is (almost) independent of {the} time
\cite{Yak2011} (being just a number that reflects superfluid
properties of a cooling neutron star). If $f_\ell$ is evaluated for
a number of neutron stars, one can generally compare their
superfluid properties on the same footing (regardless {of} their
ages) which is a great advantage of this method.

In order to use the method and determine the internal temperature of
a star $\Tg$ from the surface temperature $T_{s}^\infty$, we need to
specify {the} mass of light elements (hydrogen, helium and carbon)
in the heat blanketing envelope of the star \cite{PCY1997}. The
matter composed of lighter elements has {a} larger thermal
conductivity (is more heat transparent) which has to be taken into
account. Since we would like to account for the effects of {the}
surface magnetic fields $B$ of neutron stars, we employ the $\Tg -
T_s$ relation calculated in Ref.\ \cite{Pot2003} for neutron stars
with dipole magnetic fields in the heat blanketing envelopes. In
such a case, for a given internal temperature, the surface
temperature varies along the neutron star surface, and $T_s^\infty$
means a properly averaged surface temperature that determines the
photon thermal luminosity of the star.

The results for nine neutron stars (Table \ref{tab:obs}) are plotted
in  Fig.\ \ref{fig:AFigura}. Each panel corresponds to one of the
selected stars and shows allowable values of $f_\ell$ vs
(unknown) values of $\Delta M_{\rm acc}$. The horizontal thin
hatched lines ($\log f_\ell=0$) refer to standard candles (to guide
the eye). The hatched regions $f_\ell \lesssim 0.01$ and $f_\ell
\gtrsim 100$ have to be excluded because they seem unrealistic from
{the} theoretical point of view (see above). The regions $\Delta
M_{\rm acc} \gtrsim 10^{-7}\,M_\odot$ are also not realistic
\cite{PCY1997}; the mass of light elements $\Delta M_{\rm acc}$
cannot be too large (otherwise light elements at the bottom of the
heat blanketing envelope will transform into heavier ones under the
effect of electron captures and pycnonuclear reactions).

In order to plot Fig.\ \ref{fig:AFigura}, the values of $M$, $R$,
$T_s^\infty$, $t$ and $B$ have been taken from Table \ref{tab:obs}.
The grayed regions in the allowable range $0.01 \leq f_\ell \leq
100$ are mainly limited by {the} two thick solid lines calculated
including the effects of the magnetic fields. {The} upper line
corresponds to the maximum $T_s^\infty$ and $t$ from Table
\ref{tab:obs}, while {the} lower line is for the minimal
$T_s^\infty$ and $t$. In {the} cases when {the} upper line goes much
higher than {the} realistic values of $f_\ell$, it is not plotted,
and the grayed regions are limited by $f_\ell=100$. The thick
long-dashed lines are the same as the solid lines but neglecting the
effects of the magnetic fields. One can see that the effects are not
monotonic and not too strong. These conclusions naturally follow
from the results of Ref.\ \cite{Pot2003}. As explained there, the
nonmonotonic effects come from the competition of {the} classical
and quantum-mechanical effects of the magnetic field on the thermal
conductivity in the heat blanketing envelope of a star.

Let us analyze briefly the results of Fig.\ \ref{fig:AFigura}.
First, the mass $\Delta M_{\rm acc}$ of light elements in the
surface layer of a star is very important for inferring $f_\ell$
\cite{Weisskopf_etal11}. The higher $\Delta M_{\rm acc}$, the higher
$f_\ell$ that is required for the same surface temperature and age.
Depending on {the} (generally unknown) $\Delta M_{\rm acc}$, the
same star can be {a} stronger ($f_\ell>1$)  or weaker ($f_\ell<1$)
neutrino emitter than the standard neutrino candle which would lead
to {diverging} conclusions on its superfluid properties.

Among the nine selected stars, the Crab pulsar is less restrictive
to this analysis (Fig.\ \ref{fig:AFigura}). It is young; its thermal
surface emission is hidden by a surrounding nebula so that only the
upper limit on $T_s^\infty$ has been established
\cite{Weisskopf_etal11}. The solid and long-dashed lines correspond
to this upper limit. These results allow the pulsar to have actually
any amount of light elements in the surface layers except for a
rather large $\Delta M_{\rm acc}\gtrsim 10^{-10}\,M_\odot$ at a
small neutrino cooling function $f_l \lesssim 0.1$. These results
are in line with those obtained previously in Ref.\
\cite{Weisskopf_etal11}.

For {the} five other sources (J1357, Vela, B1706, B2334, 3C\ 58) we
obtain $f_\ell > 1$, i.e. {a} stronger neutrino emission than for
the case of a standard candle. This situation can occur in the
presence of {the} triplet-state pairing of neutrons in the cores of
the five neutron stars. All these stars cannot have {a} sufficiently
large $\Delta M_{\rm acc}$ (otherwise they would have been too
cold). The most restrictive among them is the Vela pulsar. It is so
cold that its neutrino cooling function should be close to the
maximum theoretical limit, and the amount of light elements in
Vela's envelope has to be small. Similar conclusions have been
mentioned in Ref.\ \cite{2016MBCAMK}.

The other three neutron stars (J1119, Pup~A and J0538), depending on
$\Delta M_{\rm acc}$, can have either $f_\ell <1$ (which can be
explained by rather strong proton superfluidity in their cores) or
$f_\ell >1$ (which is typically associated with {the} triplet-state
neutron pairing in the bulk of the core).

It is natural to assume that all neutron stars have the same EOS in
the core and the same critical temperature profiles, $T_{cn}(\rho)$
and $T_{cp}(\rho)$, for the onset of neutron and proton
superfluidities there. Certainly, they can have different masses,
radii, $\Delta M_{\rm acc}$, and magnetic fields. Changing $M$, we
vary the widths of the neutron and proton superfluidity layers, and,
therefore, $L_\nu^\infty(\Tg)$, $C_{\rm core}(\Tg)$ and $f_\ell$.
Changing $\Delta M_{\rm acc}$ and $B$, we modify $\Tg$ for a given
$T_s^\infty$ which also affects $f_\ell$. Therefore, one cannot
expect the same $\Delta M_{\rm acc}$ and $f_\ell$ for all neutron
stars. By studying the $\Delta M_{\rm acc}-f_\ell$ diagrams, one can
investigate statistical distributions of the important parameters of
cooling neutron stars. Nevertheless, note that according to Fig.\
\ref{fig:AFigura}, all the selected neutron stars are allowed to
have $1 \lesssim f_l \lesssim 100$. In the latter case their
neutrino emission is stronger than that due to the MU process, and
it is consistent with that due to the triplet-state pairing of
neutrons in the core. Therefore, no DU process is required to
explain the selected cooling neutron stars which is in line with the
use of the minimal cooling paradigm. The great advantage of our
approach is that it is almost independent of the nucleon EOS in the
neutron star core and that it allows us to analyze neutron stars
with different $T_s^\infty$ on the same footing.

If we assumed the operation of the DU processes in the selected
neutron stars, our results would be less restrictive. The results
would also be different if neutron star cores contain hyperons or
exotic matter (like free quarks, or pion or kaon condensates).

\section{Conclusions}

We have calculated the neutrino luminosities $L_\nu^\infty(\Tg)$ and
{the} heat capacities $C_{\rm core}(\Tg)$ in isothermal cores of
neutron stars with redshifted internal temperatures $\Tg$ for nine
EOSs (Table \ref{tab:EOSparams}) of superdense matter composed of
neutrons, protons, electrons, and muons. We have considered three
basic cases of neutrino luminosities (Table \ref{tab:nucases}) and
five basic cases of heat capacities (Table \ref{tab:capcases}). For
any case and any EOS, we have calculated $L_\nu^\infty(\Tg)$ and
$C_{\rm core}(\Tg)$ for a wide range of masses of neutron stars. {The
results of these calculations have been approximated by the analytic
equations (\ref{eq:LC-final}) with the parameters given in Table \ref{tab:npkabcgamma}; these parameters are independent of the
specific EOS.} We have shown
that $L_\nu^\infty(\Tg)$ and $C_{\rm core}(\Tg)$ are strongly
affected by the effects of General Relativity (Sec.\
\ref{sec:Redshift}).

{Although our analytic approximations are formally independent
of EOS in neutron star cores, they are certainly not exact and
could be improved in the future. Let us warn that they are
obtained for nucleon matter in neutron star cores and cannot
be used for the cores with a more complicated composition (containing,
for instance, hyperons). Even for the nucleon cores, the approximations
are based on a restricted grid of the EOSs and we cannot guarantee 
that they are sufficiently accurate for many other EOSs. Moreover,
our calculations are done using fixed effective masses of nucleons
and model expressions for the neutrino emissivities and specific
heat capacities. More advanced and reliable calculations of these quantities
could change the integrated neutrino luminosities and heat capacities.
It is possible that the updated approximations of
$L_\nu^\infty$ and $C_{\rm core}$ will have the same form (\ref{eq:LC-final}) 
but with somewhat different fit parameters. The fit parameters,
improved in this way, may depend on specific EOS, which might be
taken into account by choosing different parameters (e.g., $a_1,\dots,
a_5$) for different EOSs, but this is the problem for the future. 
We think, that even now, by a careful analysis of realistic uncertainities
introduced by the effects of nuclear physics, 
one can estimate allowable variations of fit parameters under these effects,
but such a complicated problems goes out of the scope of this paper.}

For illustration, we have used the approximated $L_\nu^\infty(\Tg)$
and $C_{\rm core}(\Tg)$ to analyze the most important neutrino
cooling functions $f_\ell$, Eq.\ (\ref{eq:fl}), of nine selected
isolated neutron stars (Table \ref{tab:obs}) from observations of
their thermal surface emission. We have adopted the minimal cooling
paradigm \cite{PageMinCool2004, 2004Gusakov} and have determined
{the} allowable ranges of $f_\ell$ (Fig.\ \ref{fig:AFigura}) of
these stars as functions of the mass $\Delta M_{\rm acc}$ of light
elements in {the} heat blanketing envelopes of neutron stars and
neutron star magnetic fields $B$. While the dependence of $f_\ell$
on $\Delta M_{\rm acc}$ is crucial, the dependence on $B$ turns out
to be less important. We have found that {the} typical values of
$f_\ell$ for the majority of these stars are higher than 1 (that is
higher than for the standard neutrino candle) but lower than 100
(the maximum $f_\ell$ that can be provided by the enhanced neutrino
emission due to {the} triplet-state Cooper pairing of neutrons).

Our analysis {of observations} is definitely incomplete 
in many respects. We have
considered only 9 ordinary, middle-aged, isolated neutron stars
whose thermal surface emission and age have been measured
(constrained); the total number of detected objects of this type
is {larger} than 20. All the selected stars are at the neutrino
cooling stage while some others have already passed to the photon
cooling stage. We have taken fixed values of $M$ and $R$ which were
mostly assumed to determine $T_s^\infty$ while fitting the observed
spectra with neutron star atmosphere models. It would be more
instructive to consider a grid of $M$ and $R$ for each neutron star
and use {the} obtained confidence ranges of $M$, $R$, and
$T_s^\infty$ (just as in analyzing the data on the neutron star in
HESS J1731--347 \cite{Kloch2015,Of2015-NS}). This may result in more
accurate values of $T_s^\infty$ and it may give, additionally, some
estimates on $M$ and $R$. Moreover, we could go beyond the minimal
cooling paradigm and allow for the appearance of the DU process in
massive neutron stars; we could also consider neutron star cores of
exotic composition. However, all these problems are beyond the scope
of our paper.

The results of the paper can also be used to investigate {the}
thermal
states of accreting neutron stars in XRTs (Sec.\ \ref{s:I}).\\

\begin{acknowledgements}
DDO and DGY are grateful to Peter Shternin and Valery Suleimanov for
useful advices. The work of DGY and DDO was supported by the Russian
Science Foundation (grant 14-12-00316) and the work of PH, LZ and MF
by the Polish NCN research grant no. 2013/11/B/ST9/04528. One of the
authors (DDO) is grateful to N.\ Copernicus Astronomical Center for
hospitality and perfect working conditions.
\end{acknowledgements}

\bibliographystyle{apsrev}

\begin{thebibliography}{58}
\expandafter\ifx\csname natexlab\endcsname\relax\def\natexlab#1{#1}\fi
\expandafter\ifx\csname bibnamefont\endcsname\relax
  \def\bibnamefont#1{#1}\fi
\expandafter\ifx\csname bibfnamefont\endcsname\relax
  \def\bibfnamefont#1{#1}\fi
\expandafter\ifx\csname citenamefont\endcsname\relax
  \def\citenamefont#1{#1}\fi
\expandafter\ifx\csname url\endcsname\relax
  \def\url#1{\texttt{#1}}\fi
\expandafter\ifx\csname urlprefix\endcsname\relax\def\urlprefix{URL }\fi
\providecommand{\bibinfo}[2]{#2}
\providecommand{\eprint}[2][]{\url{#2}}

\bibitem[{\citenamefont{{Yakovlev} and {Haensel}}(2003)}]{YakHaens2003}
\bibinfo{author}{\bibfnamefont{D.~G.} \bibnamefont{{Yakovlev}}}
  \bibnamefont{and}
  \bibinfo{author}{\bibfnamefont{P.}~\bibnamefont{{Haensel}}},
  \bibinfo{journal}{Astron. Astrophys.} \textbf{\bibinfo{volume}{407}},
  \bibinfo{pages}{259} (\bibinfo{year}{2003}), \eprint{astro-ph/0209026}.

\bibitem[{\citenamefont{{Yakovlev} et~al.}(2003)\citenamefont{{Yakovlev},
  {Levenfish}, and {Haensel}}}]{YLH2003}
\bibinfo{author}{\bibfnamefont{D.~G.} \bibnamefont{{Yakovlev}}},
  \bibinfo{author}{\bibfnamefont{K.~P.} \bibnamefont{{Levenfish}}},
  \bibnamefont{and}
  \bibinfo{author}{\bibfnamefont{P.}~\bibnamefont{{Haensel}}},
  \bibinfo{journal}{Astron. Astrophys.} \textbf{\bibinfo{volume}{407}},
  \bibinfo{pages}{265} (\bibinfo{year}{2003}), \eprint{astro-ph/0209027}.

\bibitem[{\citenamefont{{Yakovlev} and {Pethick}}(2004)}]{YakPeth2004}
\bibinfo{author}{\bibfnamefont{D.~G.} \bibnamefont{{Yakovlev}}}
  \bibnamefont{and} \bibinfo{author}{\bibfnamefont{C.~J.}
  \bibnamefont{{Pethick}}}, \bibinfo{journal}{Annu. Rev. Astron. Astrophys.}
  \textbf{\bibinfo{volume}{42}}, \bibinfo{pages}{169} (\bibinfo{year}{2004}).

\bibitem[{\citenamefont{{Page} et~al.}(2009)\citenamefont{{Page}, {Lattimer},
  {Prakash}, and {Steiner}}}]{PLPS2009}
\bibinfo{author}{\bibfnamefont{D.}~\bibnamefont{{Page}}},
  \bibinfo{author}{\bibfnamefont{J.~M.} \bibnamefont{{Lattimer}}},
  \bibinfo{author}{\bibfnamefont{M.}~\bibnamefont{{Prakash}}},
  \bibnamefont{and} \bibinfo{author}{\bibfnamefont{A.~W.}
  \bibnamefont{{Steiner}}}, \bibinfo{journal}{Astrophys. J.}
  \textbf{\bibinfo{volume}{707}}, \bibinfo{pages}{1131} (\bibinfo{year}{2009}),
  \eprint{0906.1621}.

\bibitem[{\citenamefont{{Potekhin} et~al.}(2015)\citenamefont{{Potekhin},
  {Pons}, and {Page}}}]{Pot2015}
\bibinfo{author}{\bibfnamefont{A.~Y.} \bibnamefont{{Potekhin}}},
  \bibinfo{author}{\bibfnamefont{J.~A.} \bibnamefont{{Pons}}},
  \bibnamefont{and} \bibinfo{author}{\bibfnamefont{D.}~\bibnamefont{{Page}}},
  \bibinfo{journal}{Space Sci. Rev.} \textbf{\bibinfo{volume}{191}},
  \bibinfo{pages}{239} (\bibinfo{year}{2015}), \eprint{1507.06186}.

\bibitem[{\citenamefont{{Yakovlev} et~al.}(2001)\citenamefont{{Yakovlev},
  {Kaminker}, {Gnedin}, and {Haensel}}}]{Yak2001}
\bibinfo{author}{\bibfnamefont{D.~G.} \bibnamefont{{Yakovlev}}},
  \bibinfo{author}{\bibfnamefont{A.~D.} \bibnamefont{{Kaminker}}},
  \bibinfo{author}{\bibfnamefont{O.~Y.} \bibnamefont{{Gnedin}}},
  \bibnamefont{and}
  \bibinfo{author}{\bibfnamefont{P.}~\bibnamefont{{Haensel}}},
  \bibinfo{journal}{Phys. Rep.} \textbf{\bibinfo{volume}{354}},
  \bibinfo{pages}{1} (\bibinfo{year}{2001}).

\bibitem[{\citenamefont{{Gudmundsson} et~al.}(1983)\citenamefont{{Gudmundsson},
  {Pethick}, and {Epstein}}}]{GPE1983}
\bibinfo{author}{\bibfnamefont{E.~H.} \bibnamefont{{Gudmundsson}}},
  \bibinfo{author}{\bibfnamefont{C.~J.} \bibnamefont{{Pethick}}},
  \bibnamefont{and} \bibinfo{author}{\bibfnamefont{R.~I.}
  \bibnamefont{{Epstein}}}, \bibinfo{journal}{Astrophys. J.}
  \textbf{\bibinfo{volume}{272}}, \bibinfo{pages}{286} (\bibinfo{year}{1983}).

\bibitem[{\citenamefont{{Haensel} et~al.}(2007)\citenamefont{{Haensel},
  {Potekhin}, and {Yakovlev}}}]{HPY2007}
\bibinfo{author}{\bibfnamefont{P.}~\bibnamefont{{Haensel}}},
  \bibinfo{author}{\bibfnamefont{A.~Y.} \bibnamefont{{Potekhin}}},
  \bibnamefont{and} \bibinfo{author}{\bibfnamefont{D.~G.}
  \bibnamefont{{Yakovlev}}}, \emph{\bibinfo{title}{{Neutron Stars. 1. Equation
  of State and Structure}}} (\bibinfo{publisher}{Springer, New York},
  \bibinfo{year}{2007}).

\bibitem[{\citenamefont{{Haensel} and {Zdunik}}(1990)}]{HZ1990}
\bibinfo{author}{\bibfnamefont{P.}~\bibnamefont{{Haensel}}} \bibnamefont{and}
  \bibinfo{author}{\bibfnamefont{J.~L.} \bibnamefont{{Zdunik}}},
  \bibinfo{journal}{Astron. Astrophys.} \textbf{\bibinfo{volume}{227}},
  \bibinfo{pages}{431} (\bibinfo{year}{1990}).

\bibitem[{\citenamefont{{Haensel} and {Zdunik}}(2003)}]{HZ2003}
\bibinfo{author}{\bibfnamefont{P.}~\bibnamefont{{Haensel}}} \bibnamefont{and}
  \bibinfo{author}{\bibfnamefont{J.~L.} \bibnamefont{{Zdunik}}},
  \bibinfo{journal}{Astron. Astrophys.} \textbf{\bibinfo{volume}{404}},
  \bibinfo{pages}{L33} (\bibinfo{year}{2003}), \eprint{astro-ph/0305220}.

\bibitem[{\citenamefont{{Haensel} and {Zdunik}}(2008)}]{HZ2008}
\bibinfo{author}{\bibfnamefont{P.}~\bibnamefont{{Haensel}}} \bibnamefont{and}
  \bibinfo{author}{\bibfnamefont{J.~L.} \bibnamefont{{Zdunik}}},
  \bibinfo{journal}{Astron. Astrophys.} \textbf{\bibinfo{volume}{480}},
  \bibinfo{pages}{459} (\bibinfo{year}{2008}), \eprint{0708.3996}.

\bibitem[{\citenamefont{{Brown} et~al.}(1998)\citenamefont{{Brown}, {Bildsten},
  and {Rutledge}}}]{Brown1998}
\bibinfo{author}{\bibfnamefont{E.~F.} \bibnamefont{{Brown}}},
  \bibinfo{author}{\bibfnamefont{L.}~\bibnamefont{{Bildsten}}},
  \bibnamefont{and} \bibinfo{author}{\bibfnamefont{R.~E.}
  \bibnamefont{{Rutledge}}}, \bibinfo{journal}{Atrophys. J. Lett.}
  \textbf{\bibinfo{volume}{504}}, \bibinfo{pages}{L95} (\bibinfo{year}{1998}),
  \eprint{astro-ph/9807179}.

\bibitem[{\citenamefont{{Cumming} et~al.}(2017)\citenamefont{{Cumming},
  {Brown}, {Fattoyev}, {Horowitz}, {Page}, and {Reddy}}}]{2017Cumming}
\bibinfo{author}{\bibfnamefont{A.}~\bibnamefont{{Cumming}}},
  \bibinfo{author}{\bibfnamefont{E.~F.} \bibnamefont{{Brown}}},
  \bibinfo{author}{\bibfnamefont{F.~J.} \bibnamefont{{Fattoyev}}},
  \bibinfo{author}{\bibfnamefont{C.~J.} \bibnamefont{{Horowitz}}},
  \bibinfo{author}{\bibfnamefont{D.}~\bibnamefont{{Page}}}, \bibnamefont{and}
  \bibinfo{author}{\bibfnamefont{S.}~\bibnamefont{{Reddy}}},
  \bibinfo{journal}{Phys. Rev. C} \textbf{\bibinfo{volume}{95}},
  \bibinfo{eid}{025806} (\bibinfo{year}{2017}), \eprint{1608.07532}.

\bibitem[{\citenamefont{{Yakovlev} et~al.}(2011)\citenamefont{{Yakovlev}, {Ho},
  {Shternin}, {Heinke}, and {Potekhin}}}]{Yak2011}
\bibinfo{author}{\bibfnamefont{D.~G.} \bibnamefont{{Yakovlev}}},
  \bibinfo{author}{\bibfnamefont{W.~C.~G.} \bibnamefont{{Ho}}},
  \bibinfo{author}{\bibfnamefont{P.~S.} \bibnamefont{{Shternin}}},
  \bibinfo{author}{\bibfnamefont{C.~O.} \bibnamefont{{Heinke}}},
  \bibnamefont{and} \bibinfo{author}{\bibfnamefont{A.~Y.}
  \bibnamefont{{Potekhin}}}, \bibinfo{journal}{MNRAS}
  \textbf{\bibinfo{volume}{411}}, \bibinfo{pages}{1977} (\bibinfo{year}{2011}).

\bibitem[{\citenamefont{{Ofengeim} et~al.}(2015)\citenamefont{{Ofengeim},
  {Kaminker}, {Klochkov}, {Suleimanov}, and {Yakovlev}}}]{Of2015-NS}
\bibinfo{author}{\bibfnamefont{D.~D.} \bibnamefont{{Ofengeim}}},
  \bibinfo{author}{\bibfnamefont{A.~D.} \bibnamefont{{Kaminker}}},
  \bibinfo{author}{\bibfnamefont{D.}~\bibnamefont{{Klochkov}}},
  \bibinfo{author}{\bibfnamefont{V.}~\bibnamefont{{Suleimanov}}},
  \bibnamefont{and} \bibinfo{author}{\bibfnamefont{D.~G.}
  \bibnamefont{{Yakovlev}}}, \bibinfo{journal}{MNRAS}
  \textbf{\bibinfo{volume}{454}}, \bibinfo{pages}{2668} (\bibinfo{year}{2015}),
  \eprint{1510.00573}.

\bibitem[{\citenamefont{{Ofengeim} et~al.}(2016)\citenamefont{{Ofengeim},
  {Fortin}, {Haensel}, {Yakovlev}, and {Zdunik}}}]{2016OFEN}
\bibinfo{author}{\bibfnamefont{D.~D.} \bibnamefont{{Ofengeim}}},
  \bibinfo{author}{\bibfnamefont{M.}~\bibnamefont{{Fortin}}},
  \bibinfo{author}{\bibfnamefont{P.}~\bibnamefont{{Haensel}}},
  \bibinfo{author}{\bibfnamefont{D.~G.} \bibnamefont{{Yakovlev}}},
  \bibnamefont{and} \bibinfo{author}{\bibfnamefont{J.~L.}
  \bibnamefont{{Zdunik}}}, \bibinfo{journal}{ArXiv e-prints}
  (\bibinfo{year}{2016}), \eprint{1612.04672}.

\bibitem[{\citenamefont{{Page} et~al.}(2015)\citenamefont{{Page}, {Lattimer},
  {Prakash}, and {Steiner}}}]{PageReview2013}
\bibinfo{author}{\bibfnamefont{D.}~\bibnamefont{{Page}}},
  \bibinfo{author}{\bibfnamefont{J.~M.} \bibnamefont{{Lattimer}}},
  \bibinfo{author}{\bibfnamefont{M.}~\bibnamefont{{Prakash}}},
  \bibnamefont{and} \bibinfo{author}{\bibfnamefont{A.~W.}
  \bibnamefont{{Steiner}}}, in \emph{\bibinfo{booktitle}{Novel Superfluids,
  vol. 2,}}, edited by \bibinfo{editor}{\bibfnamefont{K.~H.}
  \bibnamefont{{Bennemann}}} \bibnamefont{and}
  \bibinfo{editor}{\bibfnamefont{J.~B.} \bibnamefont{{Ketterson}}}
  (\bibinfo{publisher}{International Series of Monographs on Physics, vol. 157,
  505, Oxford University Press, Oxford}, \bibinfo{year}{2015}), vol.
  \bibinfo{volume}{157}, pp. \bibinfo{pages}{505--579}.

\bibitem[{\citenamefont{{Flowers} et~al.}(1976)\citenamefont{{Flowers},
  {Ruderman}, and {Sutherland}}}]{1976FRS}
\bibinfo{author}{\bibfnamefont{E.}~\bibnamefont{{Flowers}}},
  \bibinfo{author}{\bibfnamefont{M.}~\bibnamefont{{Ruderman}}},
  \bibnamefont{and}
  \bibinfo{author}{\bibfnamefont{P.}~\bibnamefont{{Sutherland}}},
  \bibinfo{journal}{Astrophys. J.} \textbf{\bibinfo{volume}{205}},
  \bibinfo{pages}{541} (\bibinfo{year}{1976}).

\bibitem[{\citenamefont{{Leinson} and {P{\'e}rez}}(2006)}]{2006LP}
\bibinfo{author}{\bibfnamefont{L.~B.} \bibnamefont{{Leinson}}}
  \bibnamefont{and}
  \bibinfo{author}{\bibfnamefont{A.}~\bibnamefont{{P{\'e}rez}}},
  \bibinfo{journal}{Physics Letters B} \textbf{\bibinfo{volume}{638}},
  \bibinfo{pages}{114} (\bibinfo{year}{2006}), \eprint{astro-ph/0606651}.

\bibitem[{\citenamefont{{Weisskopf} et~al.}(2011)\citenamefont{{Weisskopf},
  {Tennant}, {Yakovlev}, {Harding}, {Zavlin}, {O'Dell}, {Elsner}, and
  {Becker}}}]{Weisskopf_etal11}
\bibinfo{author}{\bibfnamefont{M.~C.} \bibnamefont{{Weisskopf}}},
  \bibinfo{author}{\bibfnamefont{A.~F.} \bibnamefont{{Tennant}}},
  \bibinfo{author}{\bibfnamefont{D.~G.} \bibnamefont{{Yakovlev}}},
  \bibinfo{author}{\bibfnamefont{A.}~\bibnamefont{{Harding}}},
  \bibinfo{author}{\bibfnamefont{V.~E.} \bibnamefont{{Zavlin}}},
  \bibinfo{author}{\bibfnamefont{S.~L.} \bibnamefont{{O'Dell}}},
  \bibinfo{author}{\bibfnamefont{R.~F.} \bibnamefont{{Elsner}}},
  \bibnamefont{and} \bibinfo{author}{\bibfnamefont{W.}~\bibnamefont{{Becker}}},
  \bibinfo{journal}{Astrophys. J.} \textbf{\bibinfo{volume}{743}},
  \bibinfo{eid}{139} (\bibinfo{year}{2011}).

\bibitem[{\citenamefont{{Shternin} and {Yakovlev}}(2015)}]{SY2015}
\bibinfo{author}{\bibfnamefont{P.~S.} \bibnamefont{{Shternin}}}
  \bibnamefont{and} \bibinfo{author}{\bibfnamefont{D.~G.}
  \bibnamefont{{Yakovlev}}}, \bibinfo{journal}{MNRAS}
  \textbf{\bibinfo{volume}{446}}, \bibinfo{pages}{3621} (\bibinfo{year}{2015}),
  \eprint{1411.0150}.

\bibitem[{\citenamefont{{Klochkov} et~al.}(2015)\citenamefont{{Klochkov},
  {Suleimanov}, {P{\"u}hlhofer}, {Yakovlev}, {Santangelo}, and
  {Werner}}}]{Kloch2015}
\bibinfo{author}{\bibfnamefont{D.}~\bibnamefont{{Klochkov}}},
  \bibinfo{author}{\bibfnamefont{V.}~\bibnamefont{{Suleimanov}}},
  \bibinfo{author}{\bibfnamefont{G.}~\bibnamefont{{P{\"u}hlhofer}}},
  \bibinfo{author}{\bibfnamefont{D.~G.} \bibnamefont{{Yakovlev}}},
  \bibinfo{author}{\bibfnamefont{A.}~\bibnamefont{{Santangelo}}},
  \bibnamefont{and} \bibinfo{author}{\bibfnamefont{K.}~\bibnamefont{{Werner}}},
  \bibinfo{journal}{Astron. Astrophys.} \textbf{\bibinfo{volume}{573}},
  \bibinfo{eid}{A53} (\bibinfo{year}{2015}).

\bibitem[{\citenamefont{{Lattimer} et~al.}(1991)\citenamefont{{Lattimer},
  {Pethick}, {Prakash}, and {Haensel}}}]{LPPH1991}
\bibinfo{author}{\bibfnamefont{J.~M.} \bibnamefont{{Lattimer}}},
  \bibinfo{author}{\bibfnamefont{C.~J.} \bibnamefont{{Pethick}}},
  \bibinfo{author}{\bibfnamefont{M.}~\bibnamefont{{Prakash}}},
  \bibnamefont{and}
  \bibinfo{author}{\bibfnamefont{P.}~\bibnamefont{{Haensel}}},
  \bibinfo{journal}{Physical Review Letters} \textbf{\bibinfo{volume}{66}},
  \bibinfo{pages}{2701} (\bibinfo{year}{1991}).

\bibitem[{\citenamefont{{Kaminker} et~al.}(2016)\citenamefont{{Kaminker},
  {Yakovlev}, and {Haensel}}}]{KYH2016}
\bibinfo{author}{\bibfnamefont{A.~D.} \bibnamefont{{Kaminker}}},
  \bibinfo{author}{\bibfnamefont{D.~G.} \bibnamefont{{Yakovlev}}},
  \bibnamefont{and}
  \bibinfo{author}{\bibfnamefont{P.}~\bibnamefont{{Haensel}}},
  \bibinfo{journal}{Astrophys. Sp. Sci.} \textbf{\bibinfo{volume}{361}},
  \bibinfo{eid}{267} (\bibinfo{year}{2016}), \eprint{1607.05265}.

\bibitem[{\citenamefont{{Page}}(1993)}]{Page1993}
\bibinfo{author}{\bibfnamefont{D.}~\bibnamefont{{Page}}}, in
  \emph{\bibinfo{booktitle}{Nuclear Physics in the Universe}}, edited by
  \bibinfo{editor}{\bibfnamefont{M.~W.} \bibnamefont{{Guidry}}}
  \bibnamefont{and} \bibinfo{editor}{\bibfnamefont{M.~R.}
  \bibnamefont{{Strayer}}} (\bibinfo{year}{1993}), pp.
  \bibinfo{pages}{151--162}.

\bibitem[{\citenamefont{{Fortin} et~al.}(2016)\citenamefont{{Fortin},
  {Provid{\^e}ncia}, {Raduta}, {Gulminelli}, {Zdunik}, {Haensel}, and
  {Bejger}}}]{Fortin2016}
\bibinfo{author}{\bibfnamefont{M.}~\bibnamefont{{Fortin}}},
  \bibinfo{author}{\bibfnamefont{C.}~\bibnamefont{{Provid{\^e}ncia}}},
  \bibinfo{author}{\bibfnamefont{A.~R.} \bibnamefont{{Raduta}}},
  \bibinfo{author}{\bibfnamefont{F.}~\bibnamefont{{Gulminelli}}},
  \bibinfo{author}{\bibfnamefont{J.~L.} \bibnamefont{{Zdunik}}},
  \bibinfo{author}{\bibfnamefont{P.}~\bibnamefont{{Haensel}}},
  \bibnamefont{and} \bibinfo{author}{\bibfnamefont{M.}~\bibnamefont{{Bejger}}},
  \bibinfo{journal}{\prc} \textbf{\bibinfo{volume}{94}}, \bibinfo{eid}{035804}
  (\bibinfo{year}{2016}), \eprint{1604.01944}.

\bibitem[{\citenamefont{{Douchin} and {Haensel}}(2001)}]{DH2001}
\bibinfo{author}{\bibfnamefont{F.}~\bibnamefont{{Douchin}}} \bibnamefont{and}
  \bibinfo{author}{\bibfnamefont{P.}~\bibnamefont{{Haensel}}},
  \bibinfo{journal}{Astron. Astrophys.} \textbf{\bibinfo{volume}{380}},
  \bibinfo{pages}{151} (\bibinfo{year}{2001}), \eprint{astro-ph/0111092}.

\bibitem[{\citenamefont{{Page} and {Applegate}}(1992)}]{PA1992}
\bibinfo{author}{\bibfnamefont{D.}~\bibnamefont{{Page}}} \bibnamefont{and}
  \bibinfo{author}{\bibfnamefont{J.~H.} \bibnamefont{{Applegate}}},
  \bibinfo{journal}{Astrophys. J. Lett.} \textbf{\bibinfo{volume}{394}},
  \bibinfo{pages}{L17} (\bibinfo{year}{1992}).

\bibitem[{\citenamefont{{Gusakov} et~al.}(2005)\citenamefont{{Gusakov},
  {Kaminker}, {Yakovlev}, and {Gnedin}}}]{Gus2005}
\bibinfo{author}{\bibfnamefont{M.~E.} \bibnamefont{{Gusakov}}},
  \bibinfo{author}{\bibfnamefont{A.~D.} \bibnamefont{{Kaminker}}},
  \bibinfo{author}{\bibfnamefont{D.~G.} \bibnamefont{{Yakovlev}}},
  \bibnamefont{and} \bibinfo{author}{\bibfnamefont{O.~Y.}
  \bibnamefont{{Gnedin}}}, \bibinfo{journal}{MNRAS}
  \textbf{\bibinfo{volume}{363}}, \bibinfo{pages}{555} (\bibinfo{year}{2005}).

\bibitem[{\citenamefont{{Potekhin} et~al.}(2013)\citenamefont{{Potekhin},
  {Fantina}, {Chamel}, {Pearson}, and {Goriely}}}]{BSk2013}
\bibinfo{author}{\bibfnamefont{A.~Y.} \bibnamefont{{Potekhin}}},
  \bibinfo{author}{\bibfnamefont{A.~F.} \bibnamefont{{Fantina}}},
  \bibinfo{author}{\bibfnamefont{N.}~\bibnamefont{{Chamel}}},
  \bibinfo{author}{\bibfnamefont{J.~M.} \bibnamefont{{Pearson}}},
  \bibnamefont{and}
  \bibinfo{author}{\bibfnamefont{S.}~\bibnamefont{{Goriely}}},
  \bibinfo{journal}{Astron. Astrophys.} \textbf{\bibinfo{volume}{560}},
  \bibinfo{eid}{A48} (\bibinfo{year}{2013}), \eprint{1310.0049}.

\bibitem[{\citenamefont{{Kaminker} et~al.}(2014)\citenamefont{{Kaminker},
  {Kaurov}, {Potekhin}, and {Yakovlev}}}]{KKPY14}
\bibinfo{author}{\bibfnamefont{A.~D.} \bibnamefont{{Kaminker}}},
  \bibinfo{author}{\bibfnamefont{A.~A.} \bibnamefont{{Kaurov}}},
  \bibinfo{author}{\bibfnamefont{A.~Y.} \bibnamefont{{Potekhin}}},
  \bibnamefont{and} \bibinfo{author}{\bibfnamefont{D.~G.}
  \bibnamefont{{Yakovlev}}}, \bibinfo{journal}{MNRAS}
  \textbf{\bibinfo{volume}{442}}, \bibinfo{pages}{3484} (\bibinfo{year}{2014}).

\bibitem[{\citenamefont{{Akmal} et~al.}(1998)\citenamefont{{Akmal},
  {Pandharipande}, and {Ravenhall}}}]{APR1998}
\bibinfo{author}{\bibfnamefont{A.}~\bibnamefont{{Akmal}}},
  \bibinfo{author}{\bibfnamefont{V.~R.} \bibnamefont{{Pandharipande}}},
  \bibnamefont{and} \bibinfo{author}{\bibfnamefont{D.~G.}
  \bibnamefont{{Ravenhall}}}, \bibinfo{journal}{Phys. Rev. C}
  \textbf{\bibinfo{volume}{58}}, \bibinfo{pages}{1804} (\bibinfo{year}{1998}).

\bibitem[{\citenamefont{{Demorest} et~al.}(2010)\citenamefont{{Demorest},
  {Pennucci}, {Ransom}, {Roberts}, and {Hessels}}}]{Demorest2010}
\bibinfo{author}{\bibfnamefont{P.~B.} \bibnamefont{{Demorest}}},
  \bibinfo{author}{\bibfnamefont{T.}~\bibnamefont{{Pennucci}}},
  \bibinfo{author}{\bibfnamefont{S.~M.} \bibnamefont{{Ransom}}},
  \bibinfo{author}{\bibfnamefont{M.~S.~E.} \bibnamefont{{Roberts}}},
  \bibnamefont{and} \bibinfo{author}{\bibfnamefont{J.~W.~T.}
  \bibnamefont{{Hessels}}}, \bibinfo{journal}{Nature}
  \textbf{\bibinfo{volume}{467}}, \bibinfo{pages}{1081} (\bibinfo{year}{2010}).

\bibitem[{\citenamefont{{Antoniadis} et~al.}(2013)\citenamefont{{Antoniadis},
  {Freire}, {Wex}, {Tauris}, {Lynch}, {van Kerkwijk}, and {et
  al.}}}]{Antoniadis2013}
\bibinfo{author}{\bibfnamefont{J.}~\bibnamefont{{Antoniadis}}},
  \bibinfo{author}{\bibfnamefont{P.~C.~C.} \bibnamefont{{Freire}}},
  \bibinfo{author}{\bibfnamefont{N.}~\bibnamefont{{Wex}}},
  \bibinfo{author}{\bibfnamefont{T.~M.} \bibnamefont{{Tauris}}},
  \bibinfo{author}{\bibfnamefont{R.~S.} \bibnamefont{{Lynch}}},
  \bibinfo{author}{\bibfnamefont{M.~H.} \bibnamefont{{van Kerkwijk}}},
  \bibnamefont{and} \bibinfo{author}{\bibnamefont{{et al.}}},
  \bibinfo{journal}{Science} \textbf{\bibinfo{volume}{340}},
  \bibinfo{pages}{448} (\bibinfo{year}{2013}), \eprint{1304.6875}.

\bibitem[{\citenamefont{{Fonseca} et~al.}(2016)\citenamefont{{Fonseca},
  {Pennucci}, {Ellis}, {Stairs}, {Nice}, {Ransom}, {Demorest}, {Arzoumanian},
  {Crowter}, {Dolch} et~al.}}]{Fonseca}
\bibinfo{author}{\bibfnamefont{E.}~\bibnamefont{{Fonseca}}},
  \bibinfo{author}{\bibfnamefont{T.~T.} \bibnamefont{{Pennucci}}},
  \bibinfo{author}{\bibfnamefont{J.~A.} \bibnamefont{{Ellis}}},
  \bibinfo{author}{\bibfnamefont{I.~H.} \bibnamefont{{Stairs}}},
  \bibinfo{author}{\bibfnamefont{D.~J.} \bibnamefont{{Nice}}},
  \bibinfo{author}{\bibfnamefont{S.~M.} \bibnamefont{{Ransom}}},
  \bibinfo{author}{\bibfnamefont{P.~B.} \bibnamefont{{Demorest}}},
  \bibinfo{author}{\bibfnamefont{Z.}~\bibnamefont{{Arzoumanian}}},
  \bibinfo{author}{\bibfnamefont{K.}~\bibnamefont{{Crowter}}},
  \bibinfo{author}{\bibfnamefont{T.}~\bibnamefont{{Dolch}}},
  \bibnamefont{et~al.}, \bibinfo{journal}{Astrophys. J.}
  \textbf{\bibinfo{volume}{832}}, \bibinfo{eid}{167} (\bibinfo{year}{2016}),
  \eprint{1603.00545}.

\bibitem[{\citenamefont{{Friman} and {Maxwell}}(1979)}]{FM1979}
\bibinfo{author}{\bibfnamefont{B.~L.} \bibnamefont{{Friman}}} \bibnamefont{and}
  \bibinfo{author}{\bibfnamefont{O.~V.} \bibnamefont{{Maxwell}}},
  \bibinfo{journal}{Astrophys. J.} \textbf{\bibinfo{volume}{232}},
  \bibinfo{pages}{541} (\bibinfo{year}{1979}).

\bibitem[{\citenamefont{{Shapiro} and {Teukolsky}}(1983)}]{ST1983}
\bibinfo{author}{\bibfnamefont{S.~L.} \bibnamefont{{Shapiro}}}
  \bibnamefont{and} \bibinfo{author}{\bibfnamefont{S.~A.}
  \bibnamefont{{Teukolsky}}}, \emph{\bibinfo{title}{{Black holes, white dwarfs,
  and neutron stars: The physics of compact objects}}}
  (\bibinfo{publisher}{Wiley-Interscience, New York}, \bibinfo{year}{1983}).

\bibitem[{\citenamefont{{Gnedin} et~al.}(2001)\citenamefont{{Gnedin},
  {Yakovlev}, and {Potekhin}}}]{Gned2001}
\bibinfo{author}{\bibfnamefont{O.~Y.} \bibnamefont{{Gnedin}}},
  \bibinfo{author}{\bibfnamefont{D.~G.} \bibnamefont{{Yakovlev}}},
  \bibnamefont{and} \bibinfo{author}{\bibfnamefont{A.~Y.}
  \bibnamefont{{Potekhin}}}, \bibinfo{journal}{MNRAS}
  \textbf{\bibinfo{volume}{324}}, \bibinfo{pages}{725} (\bibinfo{year}{2001}).

\bibitem[{\citenamefont{{Zdunik} and {Haensel}}(2011)}]{HZ2011}
\bibinfo{author}{\bibfnamefont{J.~L.} \bibnamefont{{Zdunik}}} \bibnamefont{and}
  \bibinfo{author}{\bibfnamefont{P.}~\bibnamefont{{Haensel}}},
  \bibinfo{journal}{Astron. Astrophys.} \textbf{\bibinfo{volume}{530}},
  \bibinfo{eid}{A137} (\bibinfo{year}{2011}), \eprint{1104.0385}.

\bibitem[{\citenamefont{{Zdunik} et~al.}(2017)\citenamefont{{Zdunik}, {Fortin},
  and {Haensel}}}]{ZH2016}
\bibinfo{author}{\bibfnamefont{J.~L.} \bibnamefont{{Zdunik}}},
  \bibinfo{author}{\bibfnamefont{M.}~\bibnamefont{{Fortin}}}, \bibnamefont{and}
  \bibinfo{author}{\bibfnamefont{P.}~\bibnamefont{{Haensel}}},
  \bibinfo{journal}{Astron. Astrophys.} \textbf{\bibinfo{volume}{599}},
  \bibinfo{eid}{A119} (\bibinfo{year}{2017}).

\bibitem[{\citenamefont{{Beznogov} and {Yakovlev}}(2015)}]{Bez2015a}
\bibinfo{author}{\bibfnamefont{M.~V.} \bibnamefont{{Beznogov}}}
  \bibnamefont{and} \bibinfo{author}{\bibfnamefont{D.~G.}
  \bibnamefont{{Yakovlev}}}, \bibinfo{journal}{MNRAS}
  \textbf{\bibinfo{volume}{452}}, \bibinfo{pages}{540} (\bibinfo{year}{2015}),
  \eprint{1507.04206}.

\bibitem[{\citenamefont{{Slane} et~al.}(2004)\citenamefont{{Slane}, {Helfand},
  {van der Swaluw}, and {Murray}}}]{SHSM04}
\bibinfo{author}{\bibfnamefont{P.}~\bibnamefont{{Slane}}},
  \bibinfo{author}{\bibfnamefont{D.~J.} \bibnamefont{{Helfand}}},
  \bibinfo{author}{\bibfnamefont{E.}~\bibnamefont{{van der Swaluw}}},
  \bibnamefont{and} \bibinfo{author}{\bibfnamefont{S.~S.}
  \bibnamefont{{Murray}}}, \bibinfo{journal}{Astrophys. J}
  \textbf{\bibinfo{volume}{616}}, \bibinfo{pages}{403} (\bibinfo{year}{2004}).

\bibitem[{\citenamefont{{Shibanov} et~al.}(2008)\citenamefont{{Shibanov},
  {Lundqvist}, {Lundqvist}, {Sollerman}, and {Zyuzin}}}]{Shibanov_etal08}
\bibinfo{author}{\bibfnamefont{Y.~A.} \bibnamefont{{Shibanov}}},
  \bibinfo{author}{\bibfnamefont{N.}~\bibnamefont{{Lundqvist}}},
  \bibinfo{author}{\bibfnamefont{P.}~\bibnamefont{{Lundqvist}}},
  \bibinfo{author}{\bibfnamefont{J.}~\bibnamefont{{Sollerman}}},
  \bibnamefont{and} \bibinfo{author}{\bibfnamefont{D.}~\bibnamefont{{Zyuzin}}},
  \bibinfo{journal}{Astron. Astrophys.} \textbf{\bibinfo{volume}{486}},
  \bibinfo{pages}{273} (\bibinfo{year}{2008}).

\bibitem[{\citenamefont{{Weisskopf} et~al.}(2004)\citenamefont{{Weisskopf},
  {O'Dell}, {Paerels}, {Elsner}, {Becker}, {Tennant}, and
  {Swartz}}}]{Weisskopf_etal04}
\bibinfo{author}{\bibfnamefont{M.~C.} \bibnamefont{{Weisskopf}}},
  \bibinfo{author}{\bibfnamefont{S.~L.} \bibnamefont{{O'Dell}}},
  \bibinfo{author}{\bibfnamefont{F.}~\bibnamefont{{Paerels}}},
  \bibinfo{author}{\bibfnamefont{R.~F.} \bibnamefont{{Elsner}}},
  \bibinfo{author}{\bibfnamefont{W.}~\bibnamefont{{Becker}}},
  \bibinfo{author}{\bibfnamefont{A.~F.} \bibnamefont{{Tennant}}},
  \bibnamefont{and} \bibinfo{author}{\bibfnamefont{D.~A.}
  \bibnamefont{{Swartz}}}, \bibinfo{journal}{Astrophys. J.}
  \textbf{\bibinfo{volume}{601}}, \bibinfo{pages}{1050} (\bibinfo{year}{2004}).

\bibitem[{\citenamefont{{Zavlin}}(2009)}]{Zavlin09}
\bibinfo{author}{\bibfnamefont{V.~E.} \bibnamefont{{Zavlin}}}, in
  \emph{\bibinfo{booktitle}{Astrophysics and Space Science Library}}, edited by
  \bibinfo{editor}{\bibfnamefont{W.}~\bibnamefont{{Becker}}}
  (\bibinfo{year}{2009}), vol. \bibinfo{volume}{357} of
  \emph{\bibinfo{series}{Astrophysics and Space Science Library}}, p.
  \bibinfo{pages}{181}.

\bibitem[{\citenamefont{{Zavlin} et~al.}(1999)\citenamefont{{Zavlin},
  {Tr{\"u}mper}, and {Pavlov}}}]{ZTP99}
\bibinfo{author}{\bibfnamefont{V.~E.} \bibnamefont{{Zavlin}}},
  \bibinfo{author}{\bibfnamefont{J.}~\bibnamefont{{Tr{\"u}mper}}},
  \bibnamefont{and} \bibinfo{author}{\bibfnamefont{G.~G.}
  \bibnamefont{{Pavlov}}}, \bibinfo{journal}{Astrophys. J.}
  \textbf{\bibinfo{volume}{525}}, \bibinfo{pages}{959} (\bibinfo{year}{1999}).

\bibitem[{\citenamefont{{Becker} et~al.}(2012)\citenamefont{{Becker}, {Prinz},
  {Winkler}, and {Petre}}}]{BPWP12}
\bibinfo{author}{\bibfnamefont{W.}~\bibnamefont{{Becker}}},
  \bibinfo{author}{\bibfnamefont{T.}~\bibnamefont{{Prinz}}},
  \bibinfo{author}{\bibfnamefont{P.~F.} \bibnamefont{{Winkler}}},
  \bibnamefont{and} \bibinfo{author}{\bibfnamefont{R.}~\bibnamefont{{Petre}}},
  \bibinfo{journal}{Astrophys. J.} \textbf{\bibinfo{volume}{755}},
  \bibinfo{eid}{141} (\bibinfo{year}{2012}).

\bibitem[{\citenamefont{{Gotthelf} and {Halpern}}(2009)}]{2009GH}
\bibinfo{author}{\bibfnamefont{E.~V.} \bibnamefont{{Gotthelf}}}
  \bibnamefont{and} \bibinfo{author}{\bibfnamefont{J.~P.}
  \bibnamefont{{Halpern}}}, \bibinfo{journal}{Astrophys. J. Lett.}
  \textbf{\bibinfo{volume}{695}}, \bibinfo{pages}{L35} (\bibinfo{year}{2009}),
  \eprint{0902.3007}.

\bibitem[{\citenamefont{{Zavlin}}(2007)}]{Zavlin07}
\bibinfo{author}{\bibfnamefont{V.~E.} \bibnamefont{{Zavlin}}},
  \bibinfo{journal}{Astrophys. J. Lett.} \textbf{\bibinfo{volume}{665}},
  \bibinfo{pages}{L143} (\bibinfo{year}{2007}).

\bibitem[{\citenamefont{{Pavlov} et~al.}(2001)\citenamefont{{Pavlov}, {Zavlin},
  {Sanwal}, {Burwitz}, and {Garmire}}}]{Pavlov_etal01}
\bibinfo{author}{\bibfnamefont{G.~G.} \bibnamefont{{Pavlov}}},
  \bibinfo{author}{\bibfnamefont{V.~E.} \bibnamefont{{Zavlin}}},
  \bibinfo{author}{\bibfnamefont{D.}~\bibnamefont{{Sanwal}}},
  \bibinfo{author}{\bibfnamefont{V.}~\bibnamefont{{Burwitz}}},
  \bibnamefont{and} \bibinfo{author}{\bibfnamefont{G.~P.}
  \bibnamefont{{Garmire}}}, \bibinfo{journal}{Astrophys. J. Lett.}
  \textbf{\bibinfo{volume}{552}}, \bibinfo{pages}{L129} (\bibinfo{year}{2001}).

\bibitem[{\citenamefont{{McGowan} et~al.}(2004)\citenamefont{{McGowan}, {Zane},
  {Cropper}, {Kennea}, {C{\'o}rdova}, {Ho}, {Sasseen}, and
  {Vestrand}}}]{McGowan_etal04}
\bibinfo{author}{\bibfnamefont{K.~E.} \bibnamefont{{McGowan}}},
  \bibinfo{author}{\bibfnamefont{S.}~\bibnamefont{{Zane}}},
  \bibinfo{author}{\bibfnamefont{M.}~\bibnamefont{{Cropper}}},
  \bibinfo{author}{\bibfnamefont{J.~A.} \bibnamefont{{Kennea}}},
  \bibinfo{author}{\bibfnamefont{F.~A.} \bibnamefont{{C{\'o}rdova}}},
  \bibinfo{author}{\bibfnamefont{C.}~\bibnamefont{{Ho}}},
  \bibinfo{author}{\bibfnamefont{T.}~\bibnamefont{{Sasseen}}},
  \bibnamefont{and} \bibinfo{author}{\bibfnamefont{W.~T.}
  \bibnamefont{{Vestrand}}}, \bibinfo{journal}{Astrophys. J.}
  \textbf{\bibinfo{volume}{600}}, \bibinfo{pages}{343} (\bibinfo{year}{2004}).

\bibitem[{\citenamefont{{Zavlin} and {Pavlov}}(2004)}]{ZP04}
\bibinfo{author}{\bibfnamefont{V.~E.} \bibnamefont{{Zavlin}}} \bibnamefont{and}
  \bibinfo{author}{\bibfnamefont{G.~G.} \bibnamefont{{Pavlov}}},
  \bibinfo{journal}{Mem. Soc. Astron. Ital.} \textbf{\bibinfo{volume}{75}},
  \bibinfo{pages}{458} (\bibinfo{year}{2004}).

\bibitem[{\citenamefont{{Manchester} et~al.}(2005)\citenamefont{{Manchester},
  {Hobbs}, {Teoh}, and {Hobbs}}}]{MANCHESTER2005}
\bibinfo{author}{\bibfnamefont{R.~N.} \bibnamefont{{Manchester}}},
  \bibinfo{author}{\bibfnamefont{G.~B.} \bibnamefont{{Hobbs}}},
  \bibinfo{author}{\bibfnamefont{A.}~\bibnamefont{{Teoh}}}, \bibnamefont{and}
  \bibinfo{author}{\bibfnamefont{M.}~\bibnamefont{{Hobbs}}},
  \bibinfo{journal}{Astron. J.} \textbf{\bibinfo{volume}{129}},
  \bibinfo{pages}{1993} (\bibinfo{year}{2005}), \eprint{astro-ph/0412641}.

\bibitem[{\citenamefont{{Page} et~al.}(2004)\citenamefont{{Page}, {Lattimer},
  {Prakash}, and {Steiner}}}]{PageMinCool2004}
\bibinfo{author}{\bibfnamefont{D.}~\bibnamefont{{Page}}},
  \bibinfo{author}{\bibfnamefont{J.~M.} \bibnamefont{{Lattimer}}},
  \bibinfo{author}{\bibfnamefont{M.}~\bibnamefont{{Prakash}}},
  \bibnamefont{and} \bibinfo{author}{\bibfnamefont{A.~W.}
  \bibnamefont{{Steiner}}}, \bibinfo{journal}{Astrophys. J. Suppl. Ser.}
  \textbf{\bibinfo{volume}{155}}, \bibinfo{pages}{623} (\bibinfo{year}{2004}),
  \eprint{astro-ph/0403657}.

\bibitem[{\citenamefont{{Gusakov} et~al.}(2004)\citenamefont{{Gusakov},
  {Kaminker}, {Yakovlev}, and {Gnedin}}}]{2004Gusakov}
\bibinfo{author}{\bibfnamefont{M.~E.} \bibnamefont{{Gusakov}}},
  \bibinfo{author}{\bibfnamefont{A.~D.} \bibnamefont{{Kaminker}}},
  \bibinfo{author}{\bibfnamefont{D.~G.} \bibnamefont{{Yakovlev}}},
  \bibnamefont{and} \bibinfo{author}{\bibfnamefont{O.~Y.}
  \bibnamefont{{Gnedin}}}, \bibinfo{journal}{Astron. Astrophys.}
  \textbf{\bibinfo{volume}{423}}, \bibinfo{pages}{1063} (\bibinfo{year}{2004}),
  \eprint{astro-ph/0404002}.

\bibitem[{\citenamefont{{Potekhin} et~al.}(1997)\citenamefont{{Potekhin},
  {Chabrier}, and {Yakovlev}}}]{PCY1997}
\bibinfo{author}{\bibfnamefont{A.~Y.} \bibnamefont{{Potekhin}}},
  \bibinfo{author}{\bibfnamefont{G.}~\bibnamefont{{Chabrier}}},
  \bibnamefont{and} \bibinfo{author}{\bibfnamefont{D.~G.}
  \bibnamefont{{Yakovlev}}}, \bibinfo{journal}{Astron. Astrophys.}
  \textbf{\bibinfo{volume}{323}}, \bibinfo{pages}{415} (\bibinfo{year}{1997}),
  \eprint{astro-ph/9706148}.

\bibitem[{\citenamefont{{Potekhin} et~al.}(2003)\citenamefont{{Potekhin},
  {Yakovlev}, {Chabrier}, and {Gnedin}}}]{Pot2003}
\bibinfo{author}{\bibfnamefont{A.~Y.} \bibnamefont{{Potekhin}}},
  \bibinfo{author}{\bibfnamefont{D.~G.} \bibnamefont{{Yakovlev}}},
  \bibinfo{author}{\bibfnamefont{G.}~\bibnamefont{{Chabrier}}},
  \bibnamefont{and} \bibinfo{author}{\bibfnamefont{O.~Y.}
  \bibnamefont{{Gnedin}}}, \bibinfo{journal}{Astrophys. J.}
  \textbf{\bibinfo{volume}{594}}, \bibinfo{pages}{404} (\bibinfo{year}{2003}),
  \eprint{astro-ph/0305256}.

\bibitem[{\citenamefont{{Beznogov} et~al.}(2016)\citenamefont{{Beznogov},
  {Fortin}, {Haensel}, {Yakovlev}, and {Zdunik}}}]{2016MBCAMK}
\bibinfo{author}{\bibfnamefont{M.~V.} \bibnamefont{{Beznogov}}},
  \bibinfo{author}{\bibfnamefont{M.}~\bibnamefont{{Fortin}}},
  \bibinfo{author}{\bibfnamefont{P.}~\bibnamefont{{Haensel}}},
  \bibinfo{author}{\bibfnamefont{D.~G.} \bibnamefont{{Yakovlev}}},
  \bibnamefont{and} \bibinfo{author}{\bibfnamefont{J.~L.}
  \bibnamefont{{Zdunik}}}, \bibinfo{journal}{MNRAS}
  \textbf{\bibinfo{volume}{463}}, \bibinfo{pages}{1307} (\bibinfo{year}{2016}),
  \eprint{1608.08091}.

\end{thebibliography}


\end{document}